\documentclass[letterpaper,11pt]{article}
\pdfoutput=1
\usepackage{jheppub_mod}
\usepackage{graphicx, amsmath, amssymb, amstext}
\usepackage{relsize, float, multirow}
\usepackage{array, epstopdf, hyperref, }
\usepackage{caption, subcaption, mathtools}

\newcommand{\sigmaz}{{\sigma^{(0)}}}
\newcommand{\sigmatwo}{{\sigma^{(2)}}}

\newcommand{\aamp}[1]{|\mathcal{M}|_{#1}}

\newcommand{\mDq}{{\mathcal{D}_q}}
\newcommand{\id}{{\rm d}}
\newcommand{\hbn}{\mathbf{\hat{n}}}
\newcommand{\uzero}{u^{(0)}}
\newcommand{\utwo}{u^{(2)}}
\newcommand{\kB}{k_{\rm B}}

\newcommand{\dfn}{\delta\! f_\nu}

\title{Dark neutrino interactions make gravitational waves blue}
\author[a]{Subhajit Ghosh,} 
\author[a]{Rishi Khatri,}
\author[a,b]{and Tuhin S.~Roy}
\affiliation[a]{Department of Theoretical Physics, Tata Institute of 
Fundamental Research, Mumbai 400005, India}
\affiliation[b]{Theory Division T-2, Los Alamos National laboratory, Los Alamos, NM 87545, USA}
\emailAdd{subhajit@theory.tifr.res.in, khatri@theory.tifr.res.in, tuhin@theory.tifr.res.in}
\date{\today}
\abstract{ 
New interactions of neutrinos can stop them from free streaming in the early Universe even after the weak decoupling epoch. This results in the enhancement of the primordial gravitational wave amplitude on small scales compared to the standard $\Lambda$CDM prediction. In this paper we calculate the effect of dark matter neutrino interactions in CMB tensor $B$-modes spectrum.  
We show that the effect of new neutrino interactions generates a scale or
$\ell$ dependent imprint in the CMB $B$-modes power spectrum at $\ell
\gtrsim 100$. In the event that primordial $B$-modes are detected by future
experiments, a departure from scale invariance, with a blue spectrum, may
not necessarily mean failure of simple inflationary models but instead may
be a sign of non-standard interactions of relativistic
particles. New interactions of neutrinos 
also induce a phase shift in the CMB B-mode power spectrum 
which cannot be mimicked by simple modifications of the primordial tensor power spectrum. There is rich information
hidden in the CMB $B$-modes spectrum beyond just the tensor to scalar ratio.
}
\preprint{TIFR/TH/17-43}
\notoc
\begin{document}
\maketitle

\section{\label{sec:intro}Introduction}
Despite steep experimental challenges, detecting the  signatures of
primordial gravitational waves (PGW) in the cosmic microwave background
(CMB) remains one of our top priorities. It is justifiably so since the
PGW, produced during a very early phase of the universe, such as
inflation, may carry signatures of physics of exceptionally high energy or
the ultraviolet (UV) physics. The evolution of the PGW and their imprint on
CMB  has been studied in
the standard $\Lambda\text{CDM}$ cosmology for many
decades~\cite{staro1979,rsv1982,polnarev1985}. A major breakthrough was the
realization that they leave a particular polarization pattern (curl modes or
the $B$-modes) in the CMB which cannot be created by the scalar modes in
linear theory~\cite{seljak1997,sz1997,kks1997}.  The CMB $B$-modes, therefore, provide a clean way of
detecting the PGW using the CMB as a detector. The results from the 
BICEP experiment~\cite{bicep2014,bicep2016}, which originally claimed to have found hints of PGW,  gave
rise to tremendous excitement in the cosmology and high energy physics community. Since then, a significant
amount of effort has gone in the direction of finding new physics that can
also source tensor perturbations~\cite{natural-inflation1,natural-inflation2,modgr1,modgr2,inflation-model1,inflation-model2,primo-mag1,primo-mag2,primo-mag3} and modify the CMB temperature
and polarization anisotropies. Naturally, this
over abundance of possible scenarios that can, in principle, give rise to the
tensor modes awakens an intellectual inverse problem  --  if and when
traces of $B$-modes are observed, how can we actually disentangle the exact
nature of UV-physics which might have caused the signal?

The purpose of this paper is not to find another UV phenomenon that might
source detectable $B$-modes. We rather point out a previously overlooked
fact that even if one predicts the exact initial conditions for the
$B$-modes, non-standard physics during the evolution through the radiation dominated
era may alter the CMB $B$-modes
spectrum non-trivially. This simply implies that the inverse problem of the
unentangling UV becomes far more challenging than we usually anticipate. In other words, new physics before the cosmological recombination may leave
a unique imprint in the primordial gravitational waves and the CMB
$B$-modes. Therefore, CMB $B$-modes, if detected, may give us information about not
only the very high energy phenomenon but also new physics operating at low
energies. In this paper we construct one proof of principle that demonstrates how new
physics before the cosmological recombination leaves characteristic
imprints in the CMB $B$-modes, providing a new probe of physics
beyond the standard model among the constituents of the universe and at the
same time confusing our view of the initial PGW. 

The physics of the evolution of the
PGW is rather simple. These are frozen in or 
conserved outside the horizon~\cite{Weinberg:2003ur}, and redshift on
subhorizon scales similar to any other radiation species with energy
density decreasing with redshift $z$ (namely,  $\rho_{\text{GW}} \propto
(1+z)^4$). The only non-trivial changes in the spectrum occur in the
presence of free streaming radiation during the evolution. We
  define a particle species to be free streaming if its mean free path  is
  greater than the horizon size. Gravitational
waves entering the horizon source anisotropic stress in the free streaming
species, which ultimately gets dissipated. The presence of the free
streaming relativistic particles,
therefore, simply results in the removal of energy from the gravitational
waves which, in turn, get damped. In terms of the Boltzmann equations for
tensor perturbations, the effect of the free streaming relativistic
particles  shows up as an
inhomogeneous piece (source or sink), which co-evolves with PGW. The
example we construct in this work relies on the fact that the inhomogeneous
piece changes anytime non-standard interactions of these radiative species
are introduced, which ultimately leaves its imprints on the tensor modes.

To be concrete, we consider neutrinos and show how their interactions can
suppress their free streaming properties. Consequently, the $B$-mode component
of the CMB escapes neutrino damping. The CMB $B$-modes, therefore, get enhanced
compared to the standard $\Lambda$CDM
solution obtained by neglecting all interactions of neutrinos.

Neutrino interactions are severely restricted when one considers the full
symmetry structure (gauged and global) of the Standard Model of particle
physics. When new degrees of freedom are included, however, non-trivial
interactions can arise even in a gauge invariant and flavor symmetric
way. There is significant cosmological evidence that new degrees of freedom
beyond the Standard Model do exist, namely dark matter(DM) and
dark energy \cite{bullet2004,bullet2004_2,planck2015}. Unfortunately, decades of WIMP (Weakly Interacting Massive Particles) searches (laboratory
experiments with atomic
nuclei~\cite{Agnese:2015nto,Akerib:2016vxi,Aprile:2017iyp,Cui:2017nnn}),
axion
searches~\cite{2010PhRvL.104d1301A,Arik:2015cjv,Vogel:2013bta,Budker:2013hfa,Abel:2017rtm},
and 
indirect searches (detection experiments looking for the annihilation or
decay of
dark matter into the Standard Model particles in $\gamma$-rays, $X$-rays as
well as the radio part of the electromagnetic
spectrum~\cite{TheFermi-LAT:2017vmf,Abdallah:2016ygi,Ahnen:2016qkx}) have
failed to yield any definitive signal for dark matter and, therefore,
failed to shine any light on the dark matter interaction with Standard
Model particles. Though Neutrino telescopes \cite{Collaboration:2011jza,Frankiewicz:2015zma,icecube2017,Primulando:2017kxf} have put constraints on DM-Neutrino interaction looking for DM annihilating to neutrinos, those bounds are highly model dependent(only applicable if DM is produced via thermal freezeout mechanism).
This
vast lack of our knowledge regarding the dark matter (especially, its
interactions with neutrinos) motivates us to go beyond simplistic scenarios
which the current experiments severely constrain and look for new probes of dark
interactions in more general settings.

 In fact, there has been a renewed interest recently in the interactions of
neutrinos with dark matter, motivated by the steady accumulation of
evidence for discrepancies between simulations that use a cold,
collisionless fluid for dark matter, and  observations at scales smaller
than galaxy clusters~\cite{Moore:1994yx,Kravtsov:1997dp}. 
 Introducing a relatively strong coupling of
dark matter to neutrinos and photon \cite{Boehm:2000gq,ah2014,kt2014,Wilkinson:2014ksa,Bertoni:2014mva,Boehm:2014vja}, with new radiative species or with itself  
\cite{abc2009,fky2010,bfm2012,cpr2014,cd2014,bsl2017,Spergel:1999mh,Foot:2014uba,Chacko:2015noa} remains one of the  simplistic ways to affect
structure formation on small scales. The interactions of the dark matter with the standard model, with other
dark particles, and with itself can influence
the cosmological observables such as  CMB scalar modes and
 the large scale structure (LSS). This work, on the other hand, implies
 that the imprints of these interactions can also be found on  CMB
 $B$-modes. Therefore, one can turn the argument around to use CMB $B$-modes as independent probes of interaction between dark matter and neutrinos.

In this work, we use a fairly generic set-up to account for the dark
matter neutrino interaction and calculate its effects on the time
evolution of tensor perturbations.  However, we emphasize that our
formalism and results are very general and apply to any new physics which
can stop the neutrinos from free streaming including neutrino self
interactions\cite{Belotsky:2001fb,Berkov:1987pz}. The only important model dependence comes from how the free
streaming properties of neutrinos evolve with the redshift. In this paper, we will
consider two different redshift dependences  motivated by particle physics. Our results apply to any new physics
which has the same time dependence as far as neutrino free streaming is
concerned and can be easily extended to models with different redshift
dependence.

This paper is organised as follows: in section~\ref{sec:model}, we begin
with the a model where we implement interactions between the dark matter
and neutrinos, calculate the size of cross-section, and its dependence on
the scale factor; in section~\ref{sec:boltz}, we derive the modifications to
the Boltzmann equations for tensor modes;  in section~\ref{sec:class}, we
summarize results of our numerical studies; and finally we conclude in
section~\ref{sec:conclusion}. We  use natural units with  the speed of
light, reduced Planck constant and Boltzmann constant $c=\hbar=k_{\rm B}=1$.

\section{\label{sec:model}Dark matter neutrino interactions as extensions of the Standard Model of particle physics}

Building models where dark matter can interact with neutrinos sufficiently
strongly so that there remain sizeable imprints on the cosmological observables
is non-trivial.  Note that we are interested in direct coupling between the
dark matter and active neutrinos, which come from the lepton electroweak
doublets. The effective interactions involving these neutrinos must,
therefore, arise in a gauge invariant way. Further, as far as the neutrino
sector is concerned, we confine ourselves entirely with the degrees of
freedom of the Standard Model(SM) of particle physics (no right-handed
neutrinos) and for the rest
of the paper we take neutrinos to be massless.
Summarizing the underlying assumptions: 

\begin{itemize}
\item In our setup, the degrees of freedom of the SM are extended to include new multiplets which are inert under the SM gauge group. 
\item The effective interaction involving neutrinos must arise from electroweak invariant operators. On one hand, it paves the way for a straightforward UV completion, and on the other, gives the natural size of the strength of interaction.  
\item Neutrinos arise from SM lepton doublets and, as a result, inherit the $SU(3)_l$ flavor symmetry, even after electroweak symmetry breaking. Additionally, we assume the neutrinos to be massless.      
\item We will focus on the dark sector that additionally preserves at least a  $U(1)_\text{D}$. 
\end{itemize}
The lowest order gauge invariant polynomial in SM fields is given by
$H^\dag l$.  This operator, is charged under global $U(1)_\text{L} \times
SU(3)_l$. We attempt to preserve the full symmetry structure in the
interaction, which implies that the polynomial with hidden sector fields
must transform under the global symmetries above. The minimum solution is
a dark-sector fermion with $-1$ lepton number and triplet under
$SU(3)_l$. Even though this would be the simplistic scenario, it does not
quite work -- after electroweak symmetry breaking these operators become
Dirac masses for SM neutrinos and we know that these Dirac adjoints can
only contribute a small fraction of dark matter density~\cite{1983ApJ...274L...1W}. Additionally, this construction does not allow for a preserved  $U(1)_\text{D}$. The minimum configuration, therefore, consists of two multiplets from the dark sector -- one candidate for dark matter (namely, $\chi$) and an additional field (denoted here by $\psi$). We will assign flavor quantum number to $\psi$, whereas  dark-charges to both $\chi$ and $\psi$. 
\begin{table}[htb!]
\label{t1}
\centering
\begin{tabular}{|c|c|c|c|c c|}
\hline
 & \multirow{2}{*}{$SU(3)_l$} & \multirow{2}{*}{$U(1)_L$} & \multirow{2}{*}{$U(1)_\text{D}$}  & \multicolumn{2}{|c|}{spin}  \\
 & & & & Case I & Case II \\
\hline 
\multirow{2}{*}{$\left( H^\dag l \right)$} & \multirow{2}{*}{$\bar{3}$} & \multirow{2}{*}{$1$} & \multirow{2}{*}{$0$} 
		& \multicolumn{2}{c|}{\multirow{2}{*}{$1/2$} }    \\
& & & &  &  \\		
\multirow{2}{*}{$\chi$} & \multirow{2}{*}{$1$} & \multirow{2}{*}{$0$} & \multirow{2}{*}{$1$} 
		& \multirow{2}{*}{$1/2$} &   \multirow{2}{*}{$0$}     \\
& & & &  &  \\		
\multirow{2}{*}{$\psi$} & \multirow{2}{*}{$3$} & \multirow{2}{*}{$-1$} & \multirow{2}{*}{$-1$} 
		& \multirow{2}{*}{$0$} &   \multirow{2}{*}{$1/2$}     \\
& & & &  &  \\		
\hline
\end{tabular}
\caption{Charge assignment of dark sector fields}
\end{table} 
Since both $\psi$, and $\chi$ carry $U(1)_\text{D}$ charges, we
additionally require $m_{\chi} \leq m_{\psi}$, for $\chi$ to be dark
matter. Note that in both cases, either $\psi$ or $\chi$ is a fermion and,
therefore, an additional spinor (the Dirac adjoint) needs to be introduced
with appropriate charges that allows us to write Dirac mass term for the
corresponding dark fermion. 

With the above charge assignments, the minimal interaction turns out to be a $5$-dimensional operator
\begin{equation}
\mathcal{L} \ \supset \  Y \frac{1}{\Lambda} \  \left( H^\dag l \right) \left( \psi \chi \right)  \qquad \Rightarrow \qquad  
	\eta \ \delta_{ij}  \ \nu_i \psi_j \chi  \quad \text{ where }  \eta \ = \   Y  \ \frac{v}{\sqrt{2}\Lambda}  \; .
\label{eq:interaction}
\end{equation}
Note that the flavor indices in $l$ and $\psi$ are contracted among each other and, therefore,  $Y$ is simply a complex number.  Also, we explicitly show flavor indices for neutrinos, to signify that neutrinos of all flavor interact with identical strength in our setup.  It is important to note that the operator $H^\dag l $, when expanded around the Higgs vacuum expectation value $v/\sqrt{2}$, yields couplings consisting of only neutrinos among all SM particles. This is particularly suitable for avoiding constraints from observables with charged leptons.

\begin{figure}[htb!]
\begin{center}
	\includegraphics[width=0.4\linewidth]{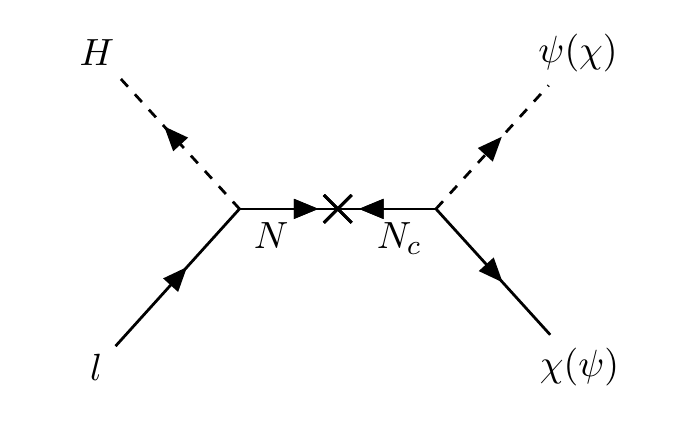}
	\caption{Feynman diagrams to demonstrate the effective operator in  Eq.~\eqref{eq:interaction} can be generated at tree level.}	
\label{fig:2}
\end{center}
\end{figure}

It is straightforward to UV complete the effective interaction shown in Eq.~\eqref{eq:interaction}. We take the opportunity to provide one such example. Consider new vector-like fermions namely $N$ with quantum numbers $\left(3, -1, 0 \right)$ under $ SU(3)_l \times U(1)_\text{L} \times U(1)_\text{D}$, and its adjoint $N^\text{c}$. The global charges allow for the following interactions 
\begin{equation}
\mathcal{L} \ \supset \  Y_N     \  N \left( H^\dag l \right)  \ + \  Y_{\bar{N}}     \  N^{\text{c}}  \left( \psi \chi \right) \ + \ M_N \ N N^{\text{c}} \, . 
\label{eq:Ltoy}
\end{equation}
At scales below $M_N$, these fermions are integrated out and one recovers the effective interactions shown in Eq.~\eqref{eq:interaction} at tree level (see Fig.~\ref{fig:2}) with the identification $Y/\Lambda = 2 \ Y_N Y_{\bar{N}} / M_N$.    
Before moving on, we emphasise that the above toy model in
Eq.~\eqref{eq:Ltoy} presents only one way to UV complete
Eq.~\eqref{eq:interaction} and the  main purpose of  Eq.~\eqref{eq:Ltoy} is to provide a proof of principle. None of the calculations shown below depend on the particularities of Eq.~\eqref{eq:Ltoy}.

\begin{figure}[htb!]
\begin{center}
	\includegraphics[width=0.8\linewidth]{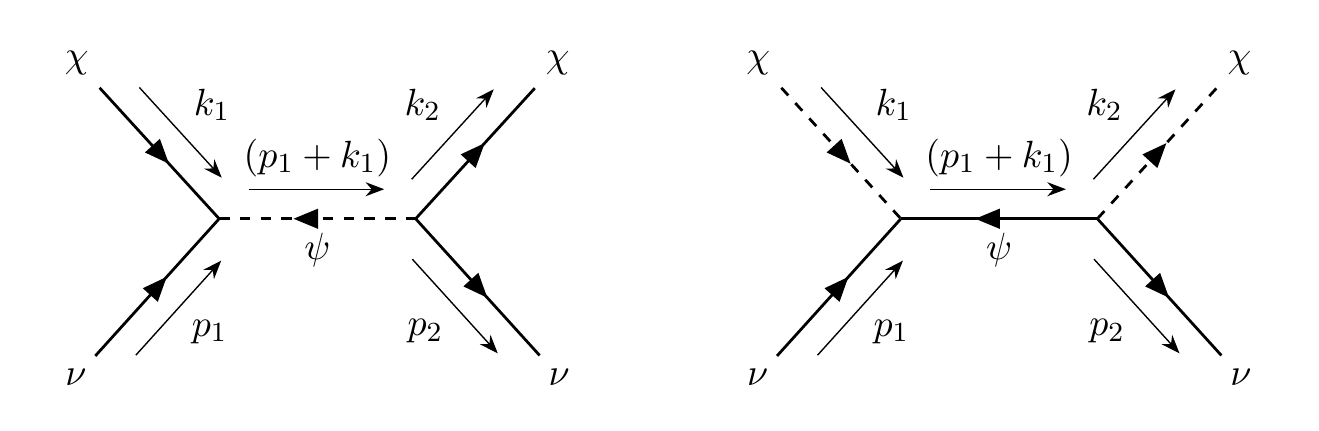}
	\caption{Feynman diagrams for dark matter neutrino scatterings. Left plot is for case I where $\chi$ is a fermion, whereas the right plot represents case II, $\chi$ being a complex scalar.}	
\label{fig:1}
\end{center}
\end{figure}

Interactions in Eq.~\eqref{eq:interaction} give rise to dark matter neutrino scattering in both cases (\textit{i.e.}, whether the dark matter $\chi$ is a fermion or a scalar). In Fig.~\ref{fig:1}, we show example Feynman diagrams that give rise to $\chi$-$\nu$ scattering.   
For massless neutrinos (which is an excellent approximation before recombination) of any flavor, the scattering matrix element between neutrinos of left helicity and dark matter (spin averaged) is simply given by 
\begin{equation}
\begin{split}
 |\mathcal{M}|^2 \ = \ 
\begin{dcases}
{2|\eta|^4 \over(-m_\chi^2\Delta + 2p_1\cdot k_1)^2 + \gamma^2}\left[k_1\cdot p_1 k_2\cdot p_2\right],\qquad 
			& \text{Case I}\\
{2|\eta|^4 \over (-m_\chi^2\Delta + 2p_1\cdot k_1)^2 +  \gamma^2 }\left[2(k_1\cdot p_1)^2 -  (p_1\cdot p_2)(m_\chi^2+2k_1\cdot p_1)\right],\qquad & \text{Case II} \,  
\end{dcases} \\
\text{where} \qquad \qquad  \Delta \ \equiv \  \frac{\left(m_\psi^2 -m_\chi^2 \right)}{m_\chi^2} \qquad 
\end{split}
\label{eq:crosssection}
\end{equation}
Here $ {p^\mu_{1,2}} $ and $ {k^\mu_{1,2}} $ are four-momentum of incoming and outgoing neutrino and $\chi$ respectively, $\gamma$ is the width of $\psi$.  As clearly seen in Fig.~\ref{fig:1}, the flavor triplet $\psi$ play the role of mediator and the cross-section depends crucially on the mass-splitting parameter of the dark sector (namely, $\Delta$).
  
We are interested in the  redshift range where the new
  interactions can stop the free streaming of neutrinos after the neutrino
  decoupling epoch ($ z\lesssim 10^9 $). Also we want to see the effects of
these new interactions on the CMB B-modes generated at the time of
recombination at $z\approx 1100$, and these are only sensitive to the evolution of
primordial gravitational waves at $z\gtrsim 1000$. In this redshift range
of interest, $10^3\lesssim z\lesssim 10^9$, massless neutrinos is an
excellent assumption and we make this assumption from now on. 
In this redshift range, we also assume that $\chi$ is non-relativistic and,
in particular, that the neutrino temperature is below DM mass (\textit{i.e.}
$\left| \bf{k}_1 \right| \ll m_\chi$).
On the other hand, throughout our redshift range of interest, neutrinos remain
relativistic with $p^2 = 0$ and follow Fermi-Dirac statistics. After
integration over the phase space, $E_\nu
\equiv \left| \bf{p} \right| $ dependence of the amplitude-squared will
give a temperature  dependent  cross section for a thermal (Fermi-Dirac) distribution of neutrinos.
We can write the general \emph{intensity averaged} cross section needed in
the Boltzmann evolution equations (see Appendix \ref{app:temp}),
$\sigma_{\chi\nu}$, as
\begin{equation}
\sigma_{\chi\nu} \ = \ \sum_{n} \ {\sigma^{(n)}
  \left(\frac{T_{\nu}}{1.95~{\rm K}}\right)^n} \, ,
\end{equation}
with one of the terms typically dominating over the rest, where 
$T_{\nu}$ is the neutrino temperature.  In $\Lambda$CDM cosmology, the
neutrino temperature is
given by $T_{\nu}=1.95(1+z)~{\rm K}$. For $n\ne 0$, $\sigma^{(n)}$ is, therefore, the
cross section extrapolated to $z=0$ assuming massless neutrinos.

We arrive at distinct scenarios depending on which term in the denominator
of Eq. \ref{eq:crosssection} dominates.  Clearly, if $\left( k_1 \cdot p_1 \right) $ dominates, the denominator goes as $E_\nu^2$, and we expect to get a temperature independent leading term. In other cases, the leading terms in the cross-section 
should go as $T_{\nu}^2$. 

\subsection*{Limit 1:  $ \Delta \ll 2\left( k_1 \cdot p_1 \right)/m_\chi^2  $ }

Note that in our redshift range of interests $E_\nu \ll
m_\chi$. Consequently, limit 1 necessarily implies $\Delta \ll 1$,
\textit{i.e.},  highly degenerate masses in the dark sector. From
a model building side, constructing a UV theory that naturally
flows towards such degenerate spectrum is non-trivial. In this work,
however, we remain agnostic of the origin of such a degeneracy and reserve any such speculations for future considerations.  
The scattering matrix element squared in limit 1 goes as, 
\begin{equation}
|\mathcal{M}|^2 \ = \
\begin{dcases}
|\eta|^4 \frac{1}{2} \big\{ 1+ \cdots. \big\} \qquad & \text{Case I}\\
|\eta|^4 \frac{1}{2} \big\{ \left(1+\hat{p}_1\cdot \hat{p}_2\right) + \cdots\big\}  \qquad & \text{Case II}
\end{dcases}
\end{equation}
The dots represents correction of $ \mathcal{O}\left(m_\chi^2 \Delta /
  \left( k_1 \cdot p_1 \right) \right)$. The leading term in the
amplitude-squared is constant as expected, which leads to a temperature
independent cross-section ($\sigma^{(0)}$). Also we define $\sigma^{(0)}$
to be the angularly averaged cross section 
 in  Case-II. The angular dependence is not important since we do not
 observe the neutrinos directly but only the average effect of their
 interactions on the gravitational waves.  In particular, we find that 
\begin{equation}
\sigma_{\chi\nu} \ \approx  \ \sigma^{(0)} \ \simeq \  10^{-13} \times \sigma_\text{Th} \times \left( \frac{\eta}{0.1}\right)^4 
	\left(\frac{m_\chi}{100~\text{GeV}} \right)^{-2}
\end{equation}  
where $ \sigma_\text{Th} = 6.65\times10^{-25}~\text{cm}^2 $ is the Thomson scattering cross-section.

\subsection*{Limit 2:  $ \Delta \gtrsim 2\left( k_1 \cdot p_1 \right)/m_\chi^2  $ }

The case with $\Delta \gg 2\left( k_1 \cdot p_1 \right)/m_\chi^2 $ is
simple to understand. In fact, in all cases where $\Delta \gtrsim 1$, we
end up with limit 2, simply because in our redshift range of interest $k_1 \cdot
p_1 \ll  m_\chi^2$. Unless special mechanisms are invoked UV models
predict splittings of order masses themselves. Even if, $\Delta$ is taken
to be negligible, renormalization drives $\Delta$ to order one values. One
can then simply replace the denominator  in Eq.~\eqref{eq:crosssection}   by $m_\chi^4 \Delta^2$, which does not have any dependence on  $E_\nu$. The leading term in the cross-section goes as $T^2$. 

Even in the case where  $\Delta \approx 2\left( k_1 \cdot p_1 \right)/m_\chi^2 $, so that the propagator falls on shell, the same temperature dependence follows. The only difference is that one simply replaces denominator by $\gamma^2$, which in these limits is given as $\gamma \sim \left( \eta^2 /16\pi^2 \right)m_\psi^2\Delta^2$. Of course, in order to write the expression of $\gamma$, we assume that $\psi$ can decay only via the interactions present in Eq.~\eqref{eq:interaction}.  Note, however, that both $\psi$ and $\chi$ carry conserved $U(1)_\text{D}$, and $\chi$ is the lightest particle with a non-zero $U(1)_\text{D}$ number. Therefore, $\chi$ should always be present in the decay product of $\psi$ irrespective of hidden sector details. Further, because of phase space considerations, we still expect $\gamma$ to vanish in the limit $\Delta \to 0$. The form of $\gamma$ is, therefore, is quite general except for the numerical pre-factor which can change from model to model. In any case, the denominators still gets replaced by a term independent of  $E_\nu$. 

Summarizing, in case $\Delta \gtrsim 1$ we obtain   
\begin{equation}\label{eq:sigmamomentum}
\aamp{}^2 \ = \ 
\begin{dcases}
2|\eta|^4 \frac{p_1^2} {m_{\chi}^2 \Delta^2 } \big\{1 + \cdots \big\}        \qquad & \text{Case I}\\
2|\eta|^4 \frac{p_1^2} {m_{\chi}^2 \Delta^2} \big\{  \left(1+\hat{p}_1\cdot \hat{p}_2 \right) + \cdots \big\}     \qquad   & \text{Case II}
\end{dcases}
\end{equation}
This ultimately yields a leading $T_\nu^2$ dependence in the cross-section
\begin{equation}
\begin{split}
\sigma_{\chi\nu} \ \approx  \ \sigma^{(2)}  \left(\frac{T_{\nu}}{1.95 ~{\rm K}}\right)^2  \ & 
\simeq \      10^{-39}\,  \sigma_\text{Th} \,  \left( \frac{T_\nu}{ 1.95~\text{K} }\right)^2  
	\left(  \frac{\eta} {0.1} \right)^4 \left( \frac{\Delta}{0.1} \right)^{-2}  \left(\frac{m_\chi}{100~\text{GeV} } \right)^{-4}   \\
&\simeq \      10^{-39}\,  \sigma_\text{Th} \,  {a}^{-2}  
\left(  \frac{\eta} {0.1} \right)^4 \left( \frac{\Delta}{0.1} \right)^{-2}
\left(\frac{m_\chi}{100~\text{GeV} } \right)^{-4}\, ,   \\
\end{split}	
\label{eq:cs-T}
\end{equation}  
where $ a$ is the scale factor normalized so that $a=1$ today.  In the last line of Eq.~\eqref{eq:cs-T} we have used the  temperature
dependence of massless neutrinos and $ T_\nu \propto 1/a $.  In the case of resonance $\Delta \approx 2\left( k_1 \cdot p_1 \right)/m_\chi^2 $, the cross-section will go as $ \sigma_{\chi\nu} \sim \Delta^{-4} $ .We note that even when we start with $\Delta
\ll 2\left( k_1 \cdot p_1 \right)/m_\chi^2$ with constant cross section,
neutrino temperature would decrease due to cosmological expansion and at
some point we will transition to limit 2: $\Delta
\gtrsim 2\left( k_1 \cdot p_1 \right)/m_\chi^2 $. In terms of dynamics, we
will, therefore, transition from a 
constant cross section  to a $T_{\nu}^2$ dependence at
some point. This will not affect our results as long as the transition
happens well after the neutrinos have decoupled from the dark matter but
may lead to a qualitative modification of dynamics if the transition
happens in the radiation dominated era when the neutrinos are still coupled
to the dark matter. We will ignore this additional complication in this paper.

\section{\label{sec:boltz}Boltzmann equations for Dark Matter - Neutrino Interaction}

At linear order in perturbation theory, tensor and scalar modes
  evolve independently\cite{doi:10.1143/PTPS.78.1}. We will only consider tensor modes in this paper
  and comment on existing results on scalar modes in section \ref{sec:class}. The non-relativistic particles like dark matter give negligible
contribution to the anisotropic stress. We, therefore, do not need to
consider the evolution of dark matter perturbations for tensor modes. Their only role,
as far as tensor modes are concerned, is to
provide scattering targets for neutrinos and stop these from free
streaming. We can also ignore the energy exchange between the
  neutrinos and the dark matter because the entropy or heat capacity of
  neutrinos is much larger than that of dark matter. To be precise, we are in the
  regime where dark matter is non-relativistic and neutrino temperature
  $T_{\nu}\ll m_{\chi}$. The energy exchange between neutrinos and dark
  matter can make the dark matter temperature equal to the neutrino
  temperature. The energy-fraction that needs to be transferred from neutrinos to
  dark matter to accomplish this is of order $\Delta \rho_{\nu}/\rho_{\nu}
  \sim n_{\chi}\kB T_{\nu}/(n_{\nu}\kB T_{\nu})= n_{\chi}/n_{\nu} \sim 10^{-9}$ for
  $m_{\chi}\sim ~{\rm GeV}$, where $n_{\chi}$ is the number density of
  interacting dark matter particles and $n_{\nu}$ is the number density of
  neutrinos. Note that this is the upper (saturation) limit to the  energy that can
  be transferred from neutrinos to dark matter at high scattering rate. In reality, since the energy
  transfer in each elastic scattering between neutrinos and dark matter is
  of order $\sim T_{\nu}/m_{\chi}\ll 1$, the actual energy lost by the
  neutrinos would become smaller as the mean free path of neutrinos
  increases with the expansion of the Universe and scattering rate
  drops. This situation is similar to the energy lost by photons to baryons
due to Compton scattering. In the photon-electron-baryon system also, because
of large entropy in photons compared to baryons (baryon to photon ratio
$\approx 6 \times 10^{-10}$), the energy lost  by photons can be neglected for most
applications except when considering small distortions of the CMB spectrum \cite{zks1968,peebles1968,cs2012,ksc2012}.

In
the following derivation we will closely follow ref~\cite{weinberg}.   The
Boltzmann equation for the neutrinos is given by
\begin{equation}
{d\!f_\nu (\mathbf{x},\mathbf{p},t) \over dt} = C[f_\nu (\mathbf{x},\mathbf{p},t)]\label{boltz}
\end{equation} 
where $ f_\nu $ is the neutrino distribution function, $ C[f_\nu] $ is
collision term arising from DM-neutrino interaction discussed in the previous section, $t$ is the proper
time, $\mathbf{x}$ is the comoving spatial coordinate and
$\mathbf{p}=p\hat{p}$ is the neutrino momentum measured by a comoving observer. We define fluctuations of neutrino distribution $ \dfn $ as
\begin{equation}
f_\nu(\mathbf{x},\mathbf{p},t) =\bar{f}_\nu + \dfn(\mathbf{x},\mathbf{p},t)
\end{equation}
where $ \bar{f}_\nu $ is the zeroth order Fermi-Dirac distribution function of the neutrinos.
Instead of $ \dfn  $, it is convenient to work with dimensionless intensity perturbation,
\begin{equation}\label{eq:defJ}
J(\mathbf{x},\hat{p},t) = {N_\nu \over a^4\bar{\rho}_\nu}\int_{0}^{\infty}\dfn(\mathbf{x},\mathbf{p},t)4\pi p^3 dp
\end{equation}
where  $ N_\nu $ is the effective number of neutrino species and $ \bar{\rho}_\nu $ is the background energy density of neutrinos.
The Boltzmann equation for tensor perturbation in terms of this new variable reads,
\begin{multline}\label{eq:j}
\frac{\partial J(\mathbf{x},\hat{p},t)}{\partial t} + \frac{\hat{p}_i}{a(t)}\frac{\partial J(\mathbf{x},\hat{p},t)}{\partial x^i} + 2\hat{p}_i\hat{p}_j{\partial \over \partial t}[D_{ij}(\mathbf{x},t)] \\ =
\begin{dcases}
-n_\chi\sigmaz\left[ J(\mathbf{x},\hat{p},t) - {1 \over 4\pi}\int d^2\hat{p'}J(\mathbf{x},\hat{p'},t)\right]& :T_\nu ~\text{independent}\\
-n_\chi\frac{\sigmatwo}{ a^{2}} \left[ J(\mathbf{x},\hat{p},t) - {1 \over 4\pi}\int d^2\hat{p'}J(\mathbf{x},\hat{p'},t)\right]& :T_\nu^2 ~\text{dependent}
\end{dcases}
\end{multline} 
with $ D_{ij} $ being the metric tensor perturbation. The detailed derivation for R.H.S of Eq~\ref{eq:j}
	is given in Appendix \ref{app:temp}. In Fourier space, we decompose $ J(\mathbf{x},\hat{p},t) $ as 
\begin{equation}
J(\mathbf{x},\hat{p},t) = \sum_{\lambda=\pm 2} \int d^3q e^{i\mathbf{q}\cdot\mathbf{x}} \beta(\mathbf{q},\lambda)e_{ij}(\hat{q},\lambda)\hat{p}_i\hat{p}_j\Delta^{T}_\nu(q,\hat{p}\cdot\hat{q},t)
\end{equation}
Here $ \beta(\mathbf{q},\lambda) $ is the  stochastic initial condition
with wave number $ \mathbf{q} $ and helicity $ \lambda $,
$e_{ij}(\hat{q},\lambda)  $ is the symmetric, divergence and traceless
polarization tensor, and $\Delta^{T}_\nu(q,\hat{p}\cdot\hat{q},t)$ is the
neutrino tensor transfer function.  One notable feature of this
decomposition is that the second integral on R.H.S of \eqref{eq:j} for both
the cases vanishes. Defining
 $\cos\theta\equiv \hat{p}\cdot\hat{q}$ we expand the neutrino transfer function in multipole components,  
\begin{equation}
\Delta^{T}_\nu(q,\cos\theta,t) = \sum_{l=0}^{\infty} {1 \over i^l}(2l+1)P_l(\cos\theta)\Delta^T_{\nu,l}(q,t)\label{eq:multipole}
\end{equation}
Following the convention of \cite{Ma:1995ey},
\begin{equation}
\delta^T_\nu \equiv \Delta^T_{\nu,0},\qquad
\theta^T_\nu \equiv {3 \over 4}q\Delta^T_{\nu,1},\qquad
\sigma^T_\nu \equiv {1 \over 2} \Delta^T_{\nu,2}
\end{equation}
and using conformal time $\eta$ defined by the relation  $\id t=a\id\eta$, we write down the modified tensor equations for neutrinos:
\begin{align}
&{\partial \delta^T_\nu \over \partial \eta } \ = \ -{4 \over 3}\theta^T_\nu - 2{\partial \mathcal{D}_q \over \partial \eta } - \dot{\mu}\delta_\nu^T\label{dis1}\\
&{\partial \theta^T_\nu \over \partial \eta } \ = \ q^2\left[{1\over 4}\delta^T_\nu - \sigma^T_\nu\right] - \dot{\mu}\theta^T_\nu\label{theta}\\
&{\partial \sigma^T_\nu \over \partial \eta } \ = \ {4 \over 15 }\theta^T_\nu - {3 \over 10}q\Delta^T_{\nu,3} - \dot{\mu} \sigma^T_\nu\label{sigma}\\
&\frac{\partial \Delta^T_{\nu,l} }{\partial \eta}  \ + \ {q \over (2l+1)}\left[(l+1)\Delta^T_{\nu,l+1} - l\Delta^T_{\nu,l-1}\right]
\ = \ -\dot{\mu}\Delta^T_{\nu,l}\label{highell}
\qquad \text{for} \quad l > 2
\end{align}
 In the above equations, $ \mathcal{D}_q $ is the Fourier transformation of tensor perturbation,
\begin{equation}
D_{ij}(\mathbf{x},t) = \sum_{\lambda=\pm 2}\int d^3q ~e^{i\mathbf{q}\cdot\mathbf{x}}~\beta(\mathbf{q},\lambda)~e_{ij}(\hat{q},\lambda)~\mathcal{D}_{q}(t)
\end{equation}
whose evolution is given by
\begin{align}
{\partial^2 \over \partial \eta^2}\mathcal{D}_q + 2aH{\partial \mathcal{D}_q \over \partial \eta} + q^2\mathcal{D}_q =16\pi a^2 G\pi^T_q\label{Eq:Dq}
\end{align}
where $ \pi^T_{q} $ is the total anisotropic stress sourced by
  photons and neutrinos and $ H~\equiv{1 \over a^2}{da \over d\eta} $ is the Hubble rate. The (dominant) neutrino part of the anisotropic stress is given by,
\begin{equation}
\pi^T_{\nu~q} = 2\bar{\rho}_\nu\left[{1 \over 15}\Delta^T_{\nu,0} + {2 \over 15}\Delta^T_{\nu,2} + {1 \over 35}\Delta^T_{\nu,4}\right]\label{piq}
\end{equation}
The last two equations are included for completeness, these remain
unaffected from the no-interaction case. In Eqs (\ref{dis1}-\ref{highell}),
$ \dot{\mu}$ is 
 the differential
\emph{optical} depth of neutrinos.    It is the inverse of comoving mean free-path of neutrino $ (\lambda_\nu = 1/\dot{\mu}) $ which is the collisional length scale of the problem. 
The model dependence of DM-neutrino interaction comes as a ratio of cross section over mass of DM. 
In general, we parametrize this interaction strength with respect to Thompson
scattering and dark matter mass $m_\chi=100~{\rm GeV}$. The expression of $ \dot{\mu} $ for the two different limits are given by
\begin{equation}
\dot{\mu} \equiv
\begin{dcases}
a \rho_\chi \sigmaz \left( 1 \over m_\chi\right) \equiv a \uzero \rho_\chi \left(\sigma_\text{th} \over 100\text{GeV}\right)\qquad &:T_{\nu}~\text{independent}\\
a\rho_\chi {\sigmatwo \over a^2} \left( 1 \over m_\chi\right) \equiv {1 \over a} \utwo \rho_\chi \left(\sigma_\text{th} \over 100\text{GeV}\right) \qquad &:T_{\nu}^2~\text{dependent}\label{Eq:mudot}
\end{dcases}
\end{equation}
where we define
\begin{equation}
 \uzero \equiv \left({\sigmaz \over \sigma_{\text{Th}}}\right)\left(100 \text{GeV} \over m_\chi \right), \qquad
 \utwo \equiv \left({\sigmatwo \over \sigma_{\text{Th}}}\right)\left(100 \text{GeV} \over m_\chi \right)
\end{equation}
and $ \rho_\chi $ is DM energy density.

\section{\label{sec:class}Results}

\begin{figure}
\resizebox{\hsize}{!}{\includegraphics{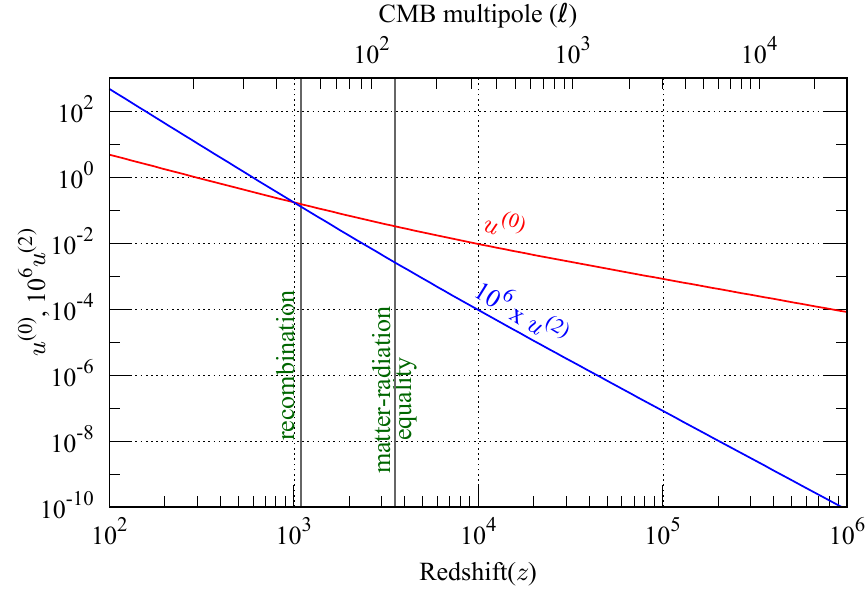}}
\caption{We have plotted the strength of neutrino dark matter interaction, $ \uzero $ (constant interaction) and $ \utwo $ (temperature-dependent interaction) as a function of redshift such that at each redshift the comoving mean free path
($\lambda_{\nu}$) of neutrino is equal to comoving Hubble radius $(1/aH)$ }.
\label{fig:mfp}
\end{figure}

\begin{figure}
\resizebox{\hsize}{!}{\includegraphics{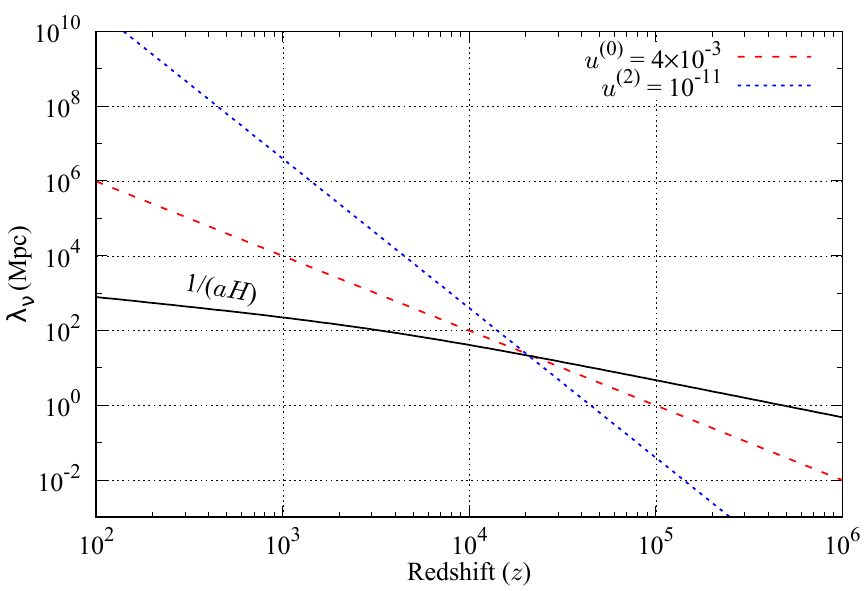}}
\caption{Evolution of the mean free path of neutrinos for our two models is
  compared with the comoving Hubble radius. The evolution  of comoving Hubble radius is shown in solid black line.}
\label{fig:mfp2}
\end{figure}

We have implemented the above mentioned modifications in the Boltzmann equation
solver CLASS (Cosmic Linear Anisotropy Solving
System)~\cite{2011JCAP...07..034B}. We only need to modify the tensor neutrino evolution equations. The evolution equation for the tensor metric perturbation and initial conditions remain unchanged. 
  We use the following flat $\Lambda$CDM cosmological parameters for all calculations~\cite{planck2015}: 
\begin{equation}
h \ = \ 0.67 \quad  \Omega_\text{b} \ = \ 0.049 \quad  \Omega_\text{cdm} = 0.26 \quad  
N_\text{eff} = 3.04  \quad  Y_{\text{He}} = 0.25  \; .
\end{equation}
with $ h = H_0/100$ where $ H_0 $ is Hubble constant, $\Omega_\text{b} $ and  $\Omega_\text{cdm}  $ are the ratio of baryon and DM energy density to critical density respectively, $ N_\text{eff} $ is the effective number of neutrinos and  $ Y_{\text{He}} $ is the helium nucleon fraction.
The amplitude and shape of CMB  $B$-modes also depends on tensor to scalar
ratio $ r $ and tensor spectral index $ n_T $, which is defined as
\begin{equation}
\mathcal{P}_T(k)=A_{T*}\left({k \over k_*}\right)^{n_T}
\end{equation}
where $ \mathcal{P}_T(k) $  is (dimensionless) primordial tensor power spectrum, $ k_* $ is pivot scale, $ A_{T*} $ is amplitude at pivot point.

 Changing $ n_T
$ tilts the power spectrum about pivot point i.e, it increases power on one side of the pivot scale w.r.t the scale invariant spectrum $ (n_T=0) $, while suppressing power on the other side. For example, setting $ n_T>0  ~(n_T<0) $ increase (decrease) power for $ k>k_* $ and decrease (increase) power for $ k<k_* $. The spectral index $ n_T $ is conventionally
fixed using the self consistency condition in terms of $ r $ assuming slow
roll inflation of single scalar field, $n_T\approx-r/8$.   However,  it can take
different values in more complicated  models of inflation. For
  the current bound on the tensor to scalar ratio, $r<0.07$
  \cite{bicep2016}, we have 
  $n_T\approx-r/8 \sim 0 $ for the fiducial  $\Lambda$CDM model with no new neutrino interactions. We use $\Lambda$CDM
	hereafter to mean the standard cosmological evolution without any new
	physics,
	\begin{equation}
	\Lambda\text{CDM Limit} \ \equiv \ \dot{\mu}\rightarrow0
	\end{equation}

\begin{figure}
\resizebox{\hsize}{!}{\includegraphics{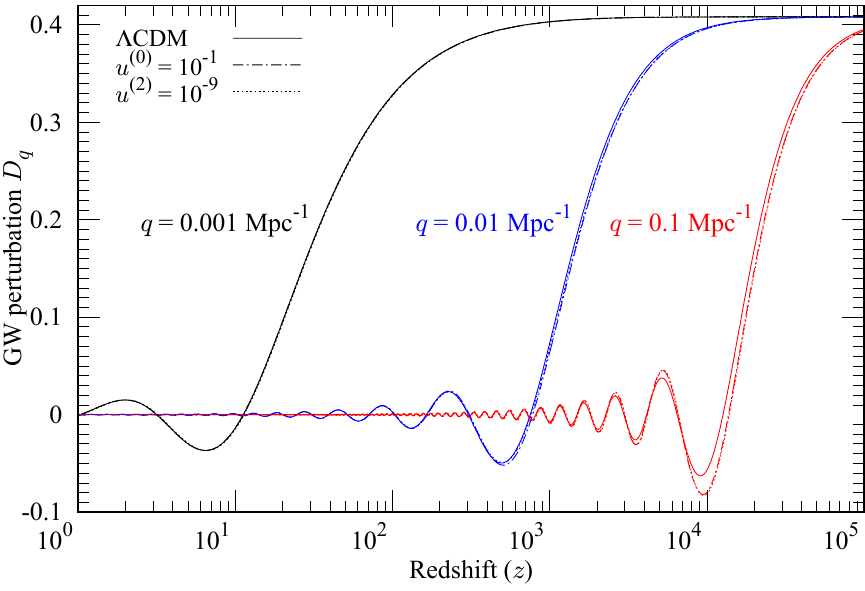}}
\caption{Evolution of the gravitational waves with three different
  wavelengths in $\Lambda$CDM cosmology and after including new neutrino
  interactions is compared. }
\label{fig:gwtransfer}
\end{figure}

Before looking at the CMB power spectra, we can try to build
  intuition about what to expect. We should expect redshifting
    due to the expansion of the Universe after a mode enters the horizon
    giving $\mathcal{D}_{q}\propto 1+z$ from second term in Eq. \ref{Eq:Dq}. There is additional damping if the neutrinos
  are free streaming when a mode enters the horizon due to anisotropic
  stress $ \pi^T_q $ in R.H.S of Eq. \ref{Eq:Dq}. This additional damping
  should be suppressed if the neutrinos are scattering with themselves or
  another particle. When  the neutrinos have a short mean streaming distance or equivalently
  large differential  optical depth $\dot{\mu}$,  the neutrino tensor
  anisotropies are suppressed.  
  This can be seen from eqns. \ref{dis1}-\ref{highell}. For very high $ \dot{\mu} $, the solution is approximately given by
\begin{equation}
\Delta_{\nu,l}^T \sim e^{-\int \dot{\mu} d\eta}
\end{equation}
Therefore,
\begin{equation}
\dot{\mu}\rightarrow \infty \qquad \Rightarrow \qquad \pi_q \rightarrow 0 
\end{equation}
 from \eqref{piq}, and the gravitational wave  (\ref{Eq:Dq}) solution deep inside the horizon becomes\cite{weinberg},
 
 \begin{equation}
 {\partial^2 \over \partial \eta^2}\mathcal{D}_q + 2aH{\partial \mathcal{D}_q \over \partial \eta} + q^2\mathcal{D}_q = 0 \qquad \Rightarrow \qquad \mathcal{D}_q \propto {1 \over q a } \sin\left(q\int d\eta\right)
 \end{equation}
In this limit gravitational
waves just oscillate with their amplitude
decreasing due to the cosmological redshifting.

A useful rough measure of  the effect on a given
  mode $q$ is given by considering whether the mean free path of the neutrinos is smaller or
  larger than the horizon scale at the time when the mode enters the
  horizon. We would expect the modes which entered the horizon when the
  mean free path of neutrinos is smaller than the horizon size would be
  amplified with respect to the $\Lambda$CDM. The modes which enter the horizon when the mean free path of the neutrinos is already larger than the horizon size
  would be damped similar to the $\Lambda$CDM case. We plot the values of
  interaction parameters $\uzero,\utwo$ as a function of redshift $z$ when
  the comoving mean free path of neutrinos is equal to the comoving Hubble radius
  $1/(aH)$ in Fig. \ref{fig:mfp}. The top axes in Fig. \ref{fig:mfp}
  shows the corresponding (approximately to a given value of
  $\uzero,\utwo$)  smallest CMB  multipole  $\ell=q(z)\left(\eta_0-\eta_{\ast}\right)$,  that would be
  affected, where $\eta_0$ is the conformal time today, $\eta_{\ast}$ is
  the conformal time at recombination,
  $\eta_0-\eta_{\ast}$ is the comoving distance to the last scattering surface and $q(z)$
is  wavenumber of  
  the mode entering the horizon at redshift $z$. For a given value of
  $\uzero,\utwo$, we can read off the multipole $\ell$ from the curves above which
  we should expect significant change in the CMB $B$-modes power
  spectrum. In Fig. \ref{fig:mfp2}, we compare the comoving mean free path
  of neutrinos with the comoving Hubble distance $1/(aH(z))$ for our two models of
  neutrino DM interaction.

\begin{figure}
\resizebox{\hsize}{!}{\includegraphics{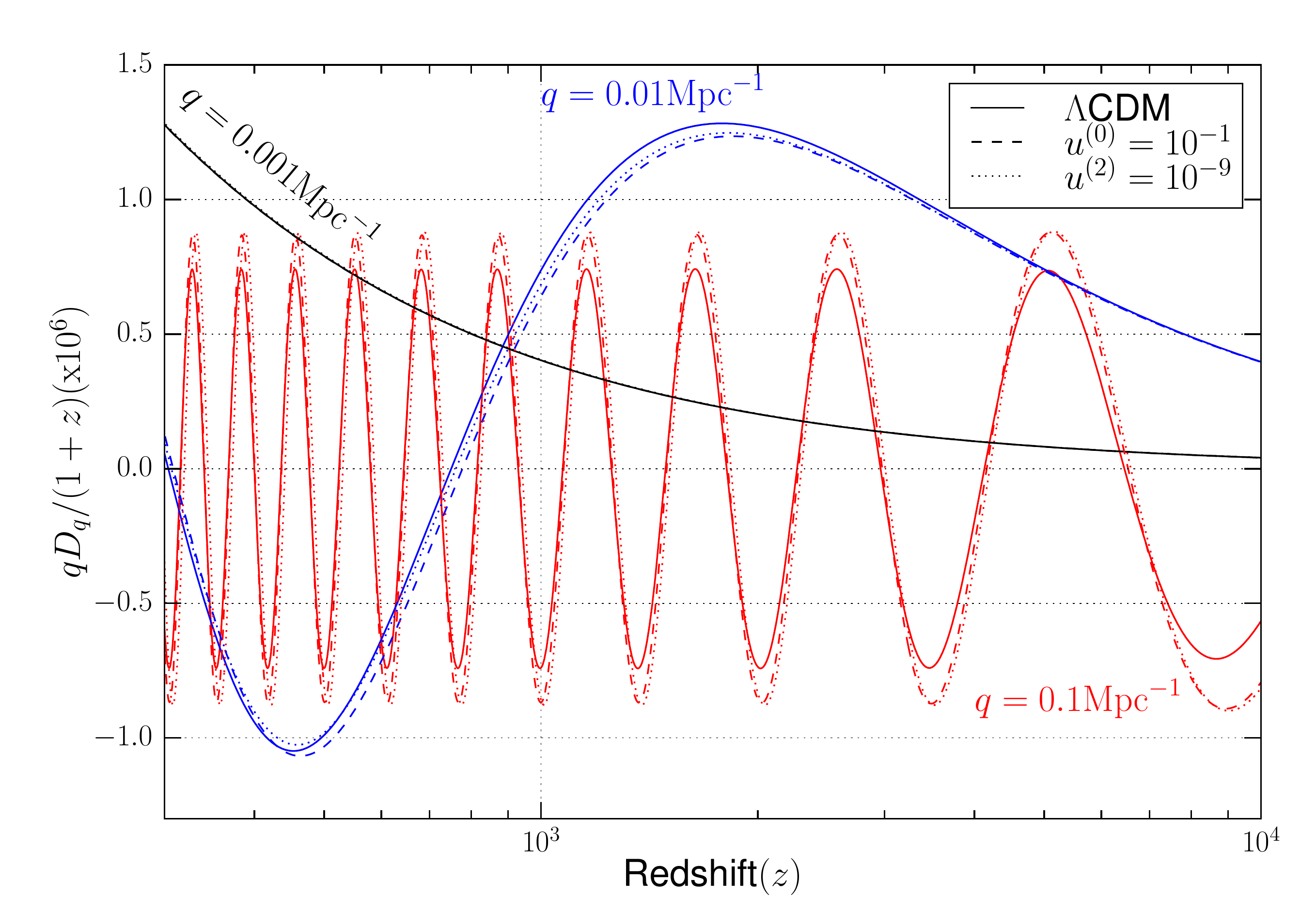}}
\caption{Zoom in version of Fig. \ref{fig:gwtransfer}.  We have divided out the
  trivial redshifting of the gravitational due to the expansion of the
  Universe and zoomed in on the redshift range near recombination.}
\label{fig:gwtransferzoom}
\end{figure}

\begin{figure}
	\resizebox{\hsize}{!}{\includegraphics{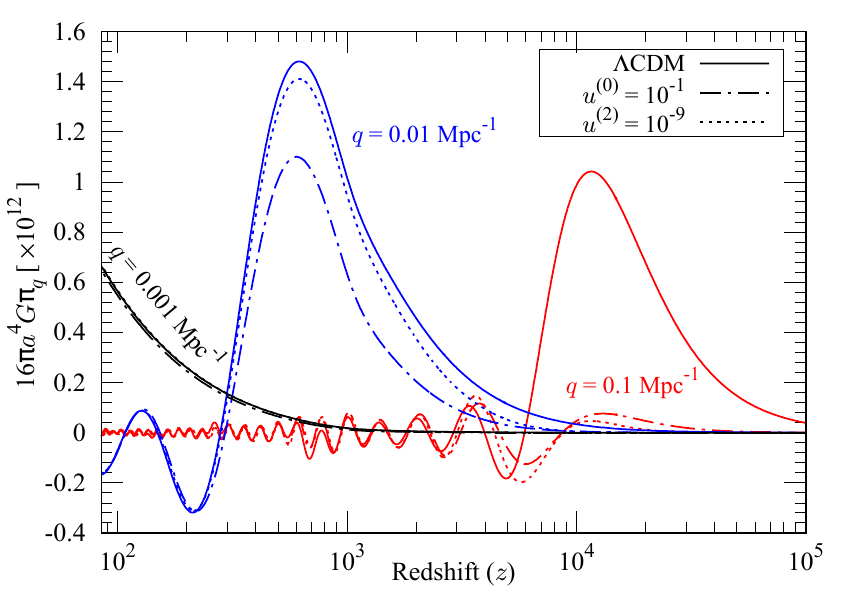}}
	\caption{Evolution of source term of Gravitational wave equation
          with three different wavelengths in $\Lambda$CDM cosmology and
          after including new neutrino interactions is
          compared.}
	\label{fig:gwsource}
\end{figure}

We show in Fig.~\ref{fig:gwtransfer}  the evolution of $\mDq$ for different
modes $q$. The curves labelled $\Lambda$CDM are the standard
evolution including both the cosmological redshift as well as the effect of
standard neutrino damping. If we turn on the neutrino interactions with a dark
particle by making the interaction parameter non-zero, the neutrino anisotropic stress
is suppressed.  We, therefore, see that the gravitational waves are enhanced
w.r.t. the $\Lambda$CDM limit for mode $q=0.1~{\rm Mpc}^{-1}$ which enters
the horizon when the neutrinos are still tightly coupled to the dark matter
and have a short mean free path. This is seen more clearly in
Fig. \ref{fig:gwtransferzoom}, where we have scaled out the cosmological
redshift of the gravitational waves. We can see in Fig. \ref{fig:mfp} that, for $\uzero=0.1$, the neutrino decoupling happens
around recombination. Therefore, for $q=0.01~{\rm Mpc}^{-1}$, which enters
the horizon just before recombination, the enhancement is less pronounced. The mode
$q=0.001~{\rm Mpc}^{-1}$ is still outside the horizon when the neutrinos
decouple  and
remains unaffected. For the case when the neutrino scattering cross section
is temperature dependent we show the evolution for $\utwo=10^{-9}$. For this
value of $\utwo$, we see from Fig. \ref{fig:mfp} that the neutrino decoupling
happens a little earlier, around matter radiation equality, giving a smaller
effect on the gravitational waves in Figs. \ref{fig:gwtransfer} and
\ref{fig:gwtransferzoom}, especially for $q=0.01~{\rm Mpc}^{-1}$, which is
just entering the horizon at this time. We also see that the
  new neutrino interactions induce a phase shift in Gravitational waves
  w.r.t. the $ \Lambda $CDM case. This phase shift is a unique signature of
  neutrino interactions and cannot be mimicked by a change in the initial
  tensor 
  power spectrum. We can see the effect of neutrino
dark matter interactions more clearly on the neutrino tensor mode
anisotropic stress plotted in Fig. \ref{fig:gwsource}. The development of
the anisotropic stress is highly suppressed for $q=0.1~{\rm Mpc}^{-1}$
mode. The suppression gets weaker as we go to smaller wavenumbers or larger
scales.

The quadrupolar anisotropy in the CMB at the last scattering surface
sourced by the gravitational waves creates polarization on Thomson
scattering, imprinting the signature of PGWs in the CMB polarization. CMB
experiments observe the $Q$ and $U$ Stokes parameters as a function of
direction in the sky (denoted here by $\hbn$), which depend on the
choice of frame. We would like to construct observables which are
independent of frame choice and can distinguish between scalar and tensor modes. Of particular interest is 
 the combination $Q\pm iU$. On rotation by angle $\alpha$,
\begin{align}
Q\pm iU \rightarrow e^{\mp 2i\alpha}\left(Q\pm iU\right)
\end{align}
and, therefore, form  spin-2 fields. They can, therefore, be decomposed into
spin-2 spherical harmonics, ${}_{\pm 2}Y_{\ell m}$\cite{Goldberg1967,sz1997},
\begin{align}
\left(Q\pm iU\right)(\hbn)=\sum_{\ell m}a_{\pm 2, \ell m}~{}_{\pm 2}Y_{\ell
  m}(\hbn)
\end{align}
We can define orthogonal combinations of spin weighted spherical harmonic coefficients $a_{\pm
  2,\ell m}$ which have a definite parity,
\begin{align}
a^E_{\ell m} \ = \ -\frac{1}{2}\left(a_{2,\ell m}+a_{-2,\ell m}\right)
\nonumber\\
a^B_{\ell m} \ = \ -i\frac{1}{2}\left(a_{2,\ell m}-a_{-2,\ell m}\right),
\end{align}
where $a^E_{\ell m}$ is a scalar (even  parity, same as the \emph{electric field}) and
$a^B_{\ell m }$ is a pseudo-scalar (odd parity, same as the \emph{magnetic
  field}). The $E$- and $B$-modes power spectra are defined by the ensemble
average (for theoretical predictions) or average over the $m$ modes (for observations),
\begin{align}
C_{\ell}^{EE}=\langle a^E_{\ell m}a^{\ast E}_{\ell m}\rangle =
\frac{1}{2\ell +1}\sum_m a^E_{\ell m}a^{\ast E}_{\ell m}\nonumber\\
C_{\ell}^{BB}=\langle a^B_{\ell m}a^{\ast B}_{\ell m}\rangle =
\frac{1}{2\ell +1}\sum_m a^B_{\ell m}a^{\ast B}_{\ell m}
\end{align}
The PGW create both $E$ and $B$ modes.
The CMB $B$-modes power spectrum, however, vanishes at linear order for scalar
perturbations (which are much larger than the tensor metric perturbations) and, therefore, provides a  powerful observable for primordial
gravitational waves. The $E$-modes created by scalar mode quadrupole at the
last scattering surface can  be converted to $B$-modes through
gravitational lensing by the large scale structure at lower redshifts. These lensing $B$-modes
provide the biggest challenge to the detection of PGW signatures in the CMB
\cite{knox2002,kesden2002,seljak2004,marian2007,smith2012,simrad2015,sherwin2015,namikawa2016,k2016,larsen2016,carron2017,sehgal2017,namikawa2017,rem2017,manzotti2017}.
We are interested in the effect of dark matter neutrino scattering on the \emph{unlensed}
$B$-modes power spectrum, $C_{\ell}^{BB}$. In the following part, we will refer
to \emph{unlensed} $B$-modes power spectrum as the \emph{$B$-modes power spectrum}.

\begin{figure}[htb!]
	\resizebox{\hsize}{!}{\includegraphics{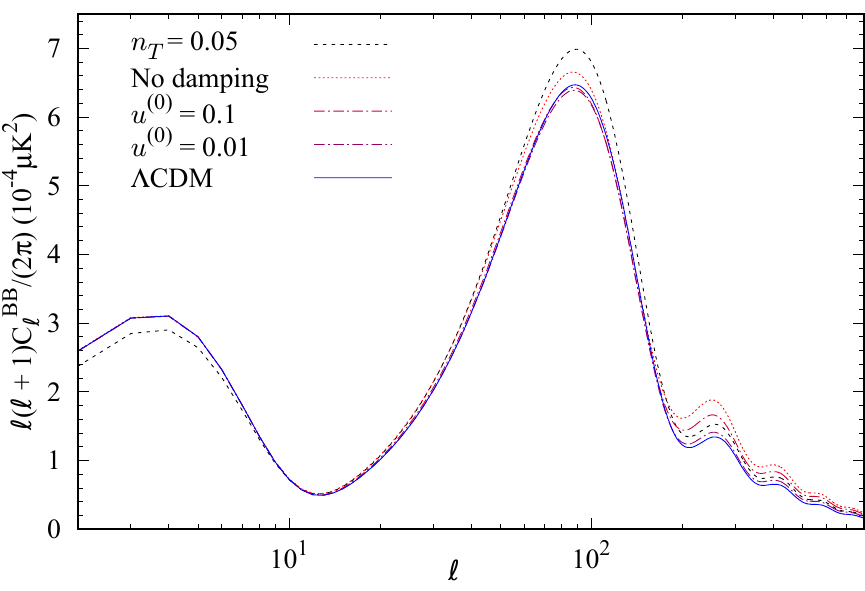}}
	\caption{Comparison of the tensor $B$-modes power spectrum of CMB for tensor to
		scalar ratio $r=0.01$ with and without neutrino dark matter
		interactions. Also shown for comparison are the power spectra for the tensor
		spectral index $n_T=0.05$ and ignoring the effect of neutrino anisotropic
		stress in gravitational waves altogether (no damping curve).}
	\label{fig:BBcompfull}
\end{figure}

\begin{figure}[htb!]
\resizebox{\hsize}{!}{\includegraphics{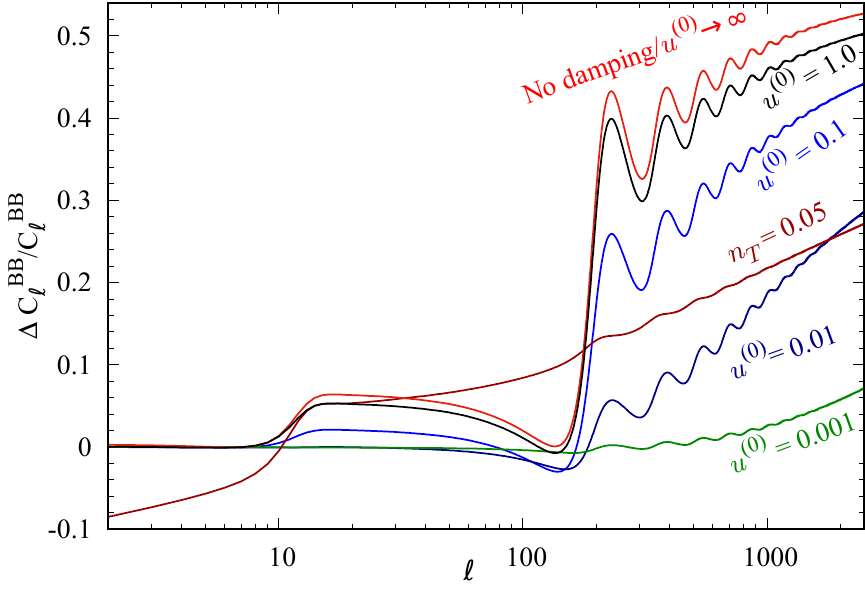}}
\caption{Fractional change of \emph{tensor} $BB$ power spectrum with different strengths
  of DM-$ \nu $ interaction compared to no interaction case for
  \emph{constant interaction strength}. We also show for comparison a blue
  initial spectrum with $n_T=0.05$. We have defined $\Delta C_l^{BB}/C_l^{BB} \equiv(C_l^{'BB}-C_l^{BB})/C_l^{BB}$, where $C_l^{'BB}$ stands for the modified $B$-modes power spectrum and  $C_l^{BB}$ stands for $\Lambda$CDM $B$-modes power spectrum.}
\label{fig:BBcomp}
\end{figure}

\begin{figure}[htb!]
\resizebox{\hsize}{!}{\includegraphics{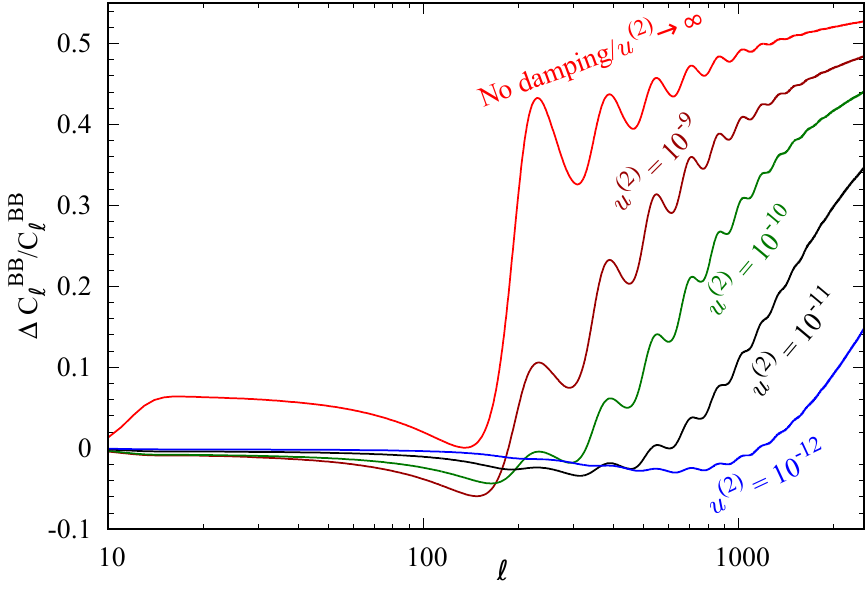}}
\caption{Fractional change of \emph{tensor} $BB$ power spectrum with different strength of DM-$ \nu $ interaction compared to no interaction case for \emph{temperature $ (T_\nu^2) $ dependent interaction strength}.}
\label{fig:bbcomp_temp}
\end{figure}

We show the effect 
 of temperature independent  DM-neutrino interaction on the CMB
$B$-modes power spectrum  in  Fig.~\ref{fig:BBcompfull}
  (full spectrum)  and Fig. \ref{fig:BBcomp} (fractional change w.r.t. $\Lambda$CDM)  and in Fig. \ref{fig:bbcomp_temp} for the $\sigma_{\chi\nu}
\propto T_{\nu}^2$ case. We also give, for reference, the CMB $B$-modes power spectrum
(labelled ``no damping'')
when we set the neutrino anisotropic stress (last term in Eq. \ref{Eq:Dq}) to
zero. This curve represents the maximum change in the CMB $B$-modes we can
expect in the extreme case when the neutrinos never decoupled and free
streamed. We see that increasing the interaction strength causes less
damping at high $ l $, and the CMB power spectrum approaches the ``no
damping'' curve as $u^{(0)},u^{(2)}\rightarrow \infty$. The phase
  shift in gravitational waves (Fig. \ref{fig:gwtransferzoom}) due to new
  neutrino interactions results in the oscillatory features in the power
  spectrum difference on top of the damping effect. We also show for reference  a blue initial tensor power
spectrum with the tensor spectral
index $n_T=0.05$ but without any new neutrino interactions.   We have chosen a  pivot scale  at $q= 0.002 $ Mpc$
^{-1} $ (pivot scale is irrelevant when $n_T\approx 0$), and tensor to
scalar ratio is $ r =0.01 $ for all curves. 

A blue initial spectrum will  also enhance the small
scale power compared to the almost scale invariant $\Lambda$CDM case for
which $n_T\approx 0$. The shape of the blue spectrum
at $\ell 
\lesssim 100$ is  different from the effect of neutrino
non-damping and sufficiently high precision measurements of the CMB
$B$-modes should be able to distinguish between the two. In particular, for a blue primordial spectrum with a constant spectral index $n_T>0$, all $\ell$ modes are affected tilting the whole $B$-modes power spectrum. In the case of interacting neutrinos  only the scales which  are either entering the horizon or are already sub-horizon at the time of recombination are enhanced w.r.t $\Lambda$CDM, while the super-horizon modes remain unaffected. We  
expect that the first detections of the CMB $B$-modes would be just above the
noise in which
case the effect of non-standard neutrino interactions may be mistaken to be
a blue initial power spectrum.

Interactions of neutrinos with dark matter would also modify the
  scalar CMB and matter power spectra. This effect has been considered
  previously in \cite{Wilkinson:2014ksa,Escudero:2015yka,dival2017}. For the scalar modes we can
  classify the changes broadly into two physical effects: 
\begin{enumerate}
\item If new interactions stop the neutrinos from free streaming, the scalar perturbations in
  them do not decay away and contribute as a source to the scalar metric
  perturbations enhancing the scalar CMB and matter power spectra.
\item If the source of the short mean free path of neutrinos is their
  scattering with dark matter particles, neutrinos provide pressure to the
  coupled neutrino-dark matter fluid, preventing the growth of dark matter
  perturbations on small scales. 
\end{enumerate}
The two main physical effects, therefore, oppose each other for the scalar
case. On the
scales below the horizon scales and above the neutrino mean free path, we have both effects. On the small scales, suppression of dark
matter and acoustic oscillations of dark matter are
the dominant effects, while on the large scales, neutrino clustering is the
dominant effect. This means that there is a range of intermediate  scales
where both  effects may be important and can partially cancel each
other.

Scalar modes (CMB as well as LSS)  are mostly sensitive to
  modifications in the dark matter power spectrum. The tensor modes, on the
  other hand, only care about what is happening to neutrinos. There is,
  therefore, a complimentarity between scalar and tensor modes. In
  particular, we can suppress the effect of neutrino
pressure on the total matter power spectrum on all scales in a multi-component dark matter
model with only a fraction of dark matter interacting with neutrinos while
keeping the effect on the tensor modes unchanged. Thus, even though there are strong
bounds when all of the dark matter interacts with neutrinos, e.g, Ref. \cite{Escudero:2015yka} puts a bound of $\uzero
\lesssim 9\times 10^{-5}$ and $\utwo
\lesssim 3\times 10^{-14}$ on the neutrino-dark matter interactions from CMB and
LSS data, they can be weakened significantly in a multi-component dark
matter model. In particular, matter power spectrum on small scales
remains unaffected if only a small fraction of dark matter interacts with
neutrinos and only the CMB constraints, which are sensitive to effects of
neutrino free streaming on large scales, remain relevant in multicomponent
dark matter models. Thus, for
example,  a
multi-component dark matter model can easily escape the much stronger
Lyman-$\alpha$ forest constraints \cite{Wilkinson:2014ksa}. 

For a
         fraction $f_{\rm i}$  of the total dark matter  interacting with
        neutrinos, we see from Eq. \ref{Eq:mudot} that the tensor mode power spectrum depends only on the
        combinations  $f_{\rm i}u^{(0)},f_{\rm i}u^{(2)}$, and we can get the
          predictions for tensor modes in multi-component models  trivially by re-labeling 
          $u^{(0)}\rightarrow f_{\rm i}u^{(0)},~u^{(2)}\rightarrow f_{\rm
              i}u^{(2)}$ in all our plots. The scalar mode total dark
            matter perturbation $\delta_{\rm DM}$ is given by 
\begin{align}
\delta_{\rm DM}=f_{\rm i}\delta_{\rm i}+\left(1-f_{\rm i}\right)\delta_{\rm CDM},
\end{align}
where $\delta_{\rm i}$ is the perturbation in the interacting component of
dark matter and $\delta_{\rm CDM}$ is the perturbation in the
non-interacting component. Only $\delta_{\rm i}$ is affected by
interactions with neutrinos, getting damped and exhibiting
oscillations because of the neutrino pressure  \cite{Wilkinson:2014ksa,Escudero:2015yka,dival2017}. As an extreme example, if we take $f_{\rm
  i}\rightarrow 0$ while keeping $f_{\rm i}u^{(0)}$ or $f_{\rm i}u^{(2)}$
constant, the effect on total dark matter perturbation vanishes while the
effect on tensor modes  remains unaffected. The only effect that 
remains on the scalar modes is the contribution of neutrinos to
the metric perturbations on large scales.

For scalar modes, a
multi-component dark matter model is studied in ~\cite{2010PhRvD..81d3507S},
where
	 they consider neutrinos interacting with a fraction of total DM
         with $ T^2 $ dependent cross section. They  parametrize the interaction by the parameter $Q$ which is related to our parameter
	$u^{(2)}$ as $Q=u^{(2)} \sigma_{\rm Th}/100 ~{\rm GeV}$.  The
        constraints from Fig. 2 of that paper show that for 	$f_{\rm
          i}\approx 10\%$ the constraints weaken considerably to  $Q\lesssim
        3 \times 10^{-40}~{\rm cm}^2~{\rm MeV}^{-1}$ or  $f_{\rm i}u^{(2)} \lesssim
        5\times 10^{-12}$.  We note that this earlier paper uses WMAP (Wilkinson Microwave Anisotropy Probe) and SDSS (Sloan Digital Sky Survey) LRG (Luminous Red Galaxies)
	data. We do not expect this result to change significantly with Planck
	since Planck only adds small scale information which remains
        unaffected when $f_{\rm i}\ll 1$. We will explore these possibilities in detail in an upcoming
paper \cite{futurepaper}.

Comparing the power spectra for the temperature independent cross-section, Fig. \ref{fig:BBcomp},  and temperature dependent cross section, Fig. \ref{fig:bbcomp_temp}, we see that, in the temperature dependent case the initial enhancement around $\ell=200$ is much smaller before it follows a similar trend as the temperature independent case at higher $\ell$. Therefore, high precision measurements of the CMB $B$-modes can in principle also tell us about not just the amplitude of the neutrino interaction cross section but also its temperature dependence.

\section{\label{sec:conclusion}Summary and conclusion}
Primordial gravitational waves are damped by the free
  streaming relativistic particles on horizon entry. In the standard
  $\Lambda$CDM cosmology, neutrinos free stream after they decouple from the
  baryonic plasma when the temperature of the Universe falls below $\sim
  1$MeV. In particular, the modes to which the CMB is sensitive, $k
\lesssim 0.1~{\rm Mpc^{-1}}$, enter the
  horizon much after neutrino decoupling and suffer damping due to neutrino
  free streaming \cite{bond1996,Weinberg:2003ur}. If  neutrinos have
  new interactions, either with dark matter or with itself, which prevents
  the neutrinos from free streaming, PGW are damped.  Compared to the standard cosmology,  we then see an amplification of the PGW amplitude
  and consequently the CMB $B$-modes for scales which enters the horizon
  when the neutrinos are still coupled. For a neutrino scattering cross
  section that is either a constant or increasing function of temperature,
  we expect the neutrinos to be more tightly coupled at higher
  redshifts. The small scale modes, which enter the horizon earlier, are therefore, more strongly
  affected compared to the long wavelength modes. In other words, the
  spectrum of gravitational waves and, therefore, the CMB $B$-modes,
  become \emph{blue} ($ n_T>0 $) compared to standard $\Lambda$CDM. 

We explore these consequences of non-standard (or dark) neutrino
  interactions within the context of a specific model of neutrino
  interacting with dark matter. We find that, as far as consequences for
  cosmology are concerned, our model allows for either a neutrino dark
  matter elastic scattering cross section $\sigma$ that is independent of
  neutrino temperature or $ \propto T_{\nu}^2$. These two cases exhaust
  almost all possibilities apart from a small portion of parameter space
  where we may have transition from a constant cross section to a
  $T_{\nu}^2$ dependent cross section. 

Quantitatively, we find that there is enhancement of the CMB
  $B$-modes at $\ell \gtrsim 100$ if neutrinos are coupled to dark matter and
  not free streaming when the modes of interest enter the horizon. This
  corresponds to  $\uzero>0.001$ for
  constant cross section case and $\utwo\gtrsim 10^{-12}$ in the case of the
  cross section having the $T_{\nu}^2$ temperature dependence (where, $ \uzero  $ and $ \utwo $ are dimensionless parameter characterising the ability of our new physics to stop the neutrino free
  streaming) assuming we
  are looking at $\ell< 1000$ modes in the CMB. In
  principle, if we could detect $\ell>1000$ modes in CMB primordial $B$-modes
  power spectrum, we could constrain even  smaller interaction strengths.

Since we enhance the small scale modes compared to the large scale modes,
the effect of new neutrino interactions mimics a blue spectrum. A blue
spectrum but with a constant spectral index $n_T$ has, however, a different CMB
$B$-modes power spectrum shape compared to the case of neutrino
interactions and can in principle be distinguished if we measure the
$B$-modes with high enough precision. To be precise, if we
  measure the B-modes power spectrum in only a few $\ell$ bins, the upward
  trend in the power spectrum due to the new neutrino interactions may be
  indistinguishable from a blue spectrum. In addition, we can mimic the
general increase in power on small scales due to new neutrino interactions
if we allow a scale dependent primordial 
spectral index, i.e. $n_T$ changing from $\sim 0$ to large positive
values as we go from large scales to small scales. However, the phase shift
in the primordial gravitational waves, which causes the oscillatory
behaviour in difference power spectrum, will be hard to mimic by simple
changes in the primordial power spectrum and provides a potential way of
distinguishing these two effects. The oscillations are, however, only a
fraction of the total change in the power spectrum and will require very precise
measurement of the B-modes power spectrum with $\ell$ resolution better than
$\delta \ell\sim 25$. We, therefore, conclude that although 
in principle it is
possible to distinguish between simple primordial tensor power spectra and 
new neutrino interactions, in practice it will be very challenging and
there is potential degeneracy between the two. This point is very
important, since in the event primordial $B$-modes are detected with a large
enough amplitude, the tensor spectral index being almost scale invariant $n_T\sim 0$
(or slightly negative),
 is expected to be a diagnostic which can rule out
a large class of inflationary models
\cite{cmbs4,core2016,prism}. Non-standard neutrino interactions confuse our
ability to test early Universe physics such as inflation.

The biggest challenge in using CMB $B$-modes as a probe of new
  physics in the neutrino/dark sector is that even though we have a
  considerably large effect ($\gtrsim 10\%$ for $u\gtrsim 10^{-3}$), most of the effect of the dark neutrino
  interactions is confined to small scales, $\ell > 100$. On these scales,
  given current limits on the tensor to scalar ratio \cite{bicep2016}, the
  CMB lensing $B$-modes, arising  from lensing of primordial E-modes by
  structure at low redshifts, dominate over the primordial $B$-modes. Our
  ability to detect the primordial $B$-modes at $\ell<100$ will, therefore,
  depend crucially on how efficiently we can delens the CMB \cite{knox2002,kesden2002,seljak2004,marian2007,smith2012,simrad2015,sherwin2015,namikawa2016,k2016,larsen2016,carron2017,sehgal2017,namikawa2017,rem2017,manzotti2017}.

Finally we note that the new interactions between neutrino and dark matter
would also affect the scalar modes and there are already constraints from
the current CMB and large scale structure data \cite{kt2014,Wilkinson:2014ksa,Bertoni:2014mva,dival2017,2010PhRvD..81d3507S,Diacoumis:2017hff,Escudero:2015yka}. The effect on
scalar modes is a sum of two effects which are opposite in sign, the
non-free streaming of neutrinos which enhances the matter perturbations and
radiation pressure in the dark matter fluid which suppresses the matter
perturbations. It may be possible to weaken these constraints by playing
these two effects against each other within some
region of parameter space while still getting large effect on the
primordial CMB $B$-modes. We will explore these possibilities in a future
publication \cite{futurepaper}.

Keeping in mind the strong focus of the CMB community on the detection of
primordial $B$-modes with ground based, balloon based and space based
experiments in the next decade, we have  tried to explore the consequences
of an eventual detection. We find that new physics in the neutrino-dark
sector may influence the cosmological signal in a non-trivial way. In
particular, in the event that the primordial $B$-modes are measured upto a
few hundred $\ell$, a deviation from a scale invariant spectrum could be
interpreted either as the failure of the simplest inflationary models or new
physics in the dark sector. These two can be disentangled if the CMB
$B$-modes are detected with high  S/N at $\ell>100$.

\section*{Acknowledgements}
TSR was supported in part by the SERB grant no ECR/2015/000196. RK  was supported
by SERB grant no. ECR/2015/000078 of Science and Engineering Research
board, Dept. of science and technology, Govt. of India. This work was
also supported by Max-Planck-Gesellschaft through the partner group
between MPI for Astrophysics, Garching and TIFR, Mumbai. 

\appendix

\section{Collision terms of DM-neutrino interaction}\label{app:temp}
In this section, we derive the collision terms for neutrino Boltzmann equations  in Section~\ref{sec:boltz}. The Boltzmann equations for tensor perturbation from Eqn \ref{boltz} reads
\begin{multline}\label{eq:coll}
{\partial\dfn(\mathbf{x},\mathbf{p},t) \over \partial t } + {p_i \over a(t)p}{\partial\dfn(\mathbf{x},\mathbf{p},t) \over \partial x^i } - {\partial \bar{f}_\nu(p) \over \partial p}{p_ip_j \over 2p}{\partial \over \partial t}D_{ij}(\mathbf{x},t)\\=
\begin{dcases}
-n_\chi\sigmaz\left[\dfn(p\hat{p}) - {1 \over 4\pi}\int d^2\hat{p'}\dfn(p\hat{p'})\right]&:\text{Limit 1}\\
-n_\chi\tilde{\sigma}{p^2 \over m_\chi^2}\left[\dfn(p\hat{p}) - {1 \over 4\pi}\int d^2\hat{p'}\dfn(p\hat{p'})\right]&:\text{Limit 2}
\end{dcases}
\end{multline}
with $ \tilde{\sigma} \equiv {1 \over 4\pi}{\eta^4 \over \Delta^2 m_{\chi}^2}$. Here we have given explicit expressions of collision terms $ C[f_\nu] $ in both the limits discussed in Section~\ref{sec:model}. After integrating over neutrino momentum we will arrive at $ T_\nu^2 $ dependent crosssection in limit 2 and then we will relate $ \tilde{\sigma} $ with $ \sigmatwo $ used in main text.
The dimensionless intensity fluctuation defined in Eq \ref{eq:defJ} is given by
\begin{equation}
J(\mathbf{x},\hat{p},t) = {N_\nu \over a^4\bar{\rho}_\nu}\int_{0}^{\infty}\dfn(\mathbf{x},\mathbf{p},t)4\pi p^3 dp
\end{equation}
\subsection{Limit 1 : ($T_\nu $ independent)}
In this limit, there is no momentum dependence in the collision term. Therefore, we can trivially carry out the momentum integration to arrive at,
\begin{multline}
\frac{\partial J(\mathbf{x},\hat{p},t)}{\partial t} + \frac{\hat{p}_i}{a(t)}\frac{\partial J(\mathbf{x},\hat{p},t)}{\partial x^i} + 2\hat{p}_i\hat{p}_j{\partial \over \partial t}[D_{ij}(\mathbf{x},t)] =
-n_\chi\sigmaz\left[ J(\mathbf{x},\hat{p},t) - {1 \over 4\pi}\int d^2\hat{p'}J(\mathbf{x},\hat{p'},t)\right]
\end{multline}
This, as anticipated, gives the \emph{$ T_\nu $-independent} collision term.
\subsection{Limit 2 : ($T^2_\nu $ dependent)}
 We can relate the fluctuations in the  neutrino distribution function to
 the  fluctuations of the neutrino temperature $\Theta$ defined by
\begin{equation}
f_\nu(\mathbf{x},p,\hat{p},t) =  \left[\exp \left( { p \over T_\nu(t)[1+\Theta(\mathbf{x},\hat{p},t)]}\right) + 1\right]^{-1}
\end{equation} 
by expanding $f_{\nu}$  to first order in $\Theta$,
\begin{equation}
\dfn(\mathbf{x},p,\hat{p},t) = -p{\partial \bar{f}_\nu \over \partial p }\Theta(\mathbf{x},\hat{p},t),
\end{equation}
where $ \bar{f}_\nu=f_\nu(\Theta=0) $ is the zeroth order Fermi-Dirac distribution for neutrino. The relation of this new variable $ \Theta(\mathbf{x},\hat{p},t) $ with $ J(\mathbf{x},\hat{p},t) $ is
\begin{align}
J(\mathbf{x},\hat{p},t)
&= {N_\nu \over a^4\bar{\rho}_\nu}\int_{0}^{\infty}\dfn(\mathbf{x},\mathbf{p},t)4\pi p^3 dp\\
&={N_\nu \over a^4\bar{\rho}_\nu} \int_{0}^{\infty}-p{\partial \bar{f}_\nu \over \partial p }\Theta(\mathbf{x},\hat{p},t)4\pi p^3 dp\\
&=4\Theta(\mathbf{x},\hat{p},t)
\end{align}
Integrating over momentum and using the relation above, we get
\begin{align}
&\frac{\partial J(\mathbf{x},\hat{p},t)}{\partial t} + \frac{\hat{p}_i}{a(t)}\frac{\partial J(\mathbf{x},\hat{p},t)}{\partial x^^i} + 2\hat{p}_i\hat{p}_j{\partial \over \partial t}[D_{ij}(\mathbf{x},t)] \\
&=-n_\chi\tilde{\sigma}{1 \over m_\chi^2}\left[ {\int_{0}^{\infty}\dfn(p\hat{p})4\pi p^5 dp \over \int_{0}^{\infty} \bar{f}_\nu 4\pi p^3 dp} - {1 \over 4\pi}\int d^2\hat{p}'{\int_{0}^{\infty}\dfn(p\hat{p'})4\pi p^5 dp \over \int_{0}^{\infty} \bar{f}_\nu 4\pi p^3 dp}  \right]\\
&=-n_\chi\tilde{\sigma}{1 \over m_\chi^2}\left[ {\int_{0}^{\infty} -p{\partial \bar{f}_\nu \over \partial p }\Theta(\hat{p}) p^5 dp \over \int_{0}^{\infty} \bar{f}_\nu p^3 dp} - {1 \over 4\pi}\int d^2\hat{p}'{\int_{0}^{\infty}-p{\partial \bar{f}_\nu \over \partial p }\Theta(\hat{p}') p^5 dp \over \int_{0}^{\infty} \bar{f}_\nu p^3 dp}  \right]\\
&=-n_\chi\tilde{\sigma}{6 \over m_\chi^2}\left(\frac{310 \pi ^2 T_\nu^2}{147}\right)\left[\Theta(\hat{p}) - {1 \over 4\pi}\int d^2\hat{p}'\Theta(\hat{p}') \right]\\
&=-n_\chi\tilde{\sigma}{6 \over 4 m_\chi^2}\left(\frac{310 \pi ^2 T_\nu^2}{147}\right)\left[J(\mathbf{x},\hat{p},t) - {1 \over 4\pi}\int d^2\hat{p}'J(\mathbf{x},\hat{p}',t) \right]\\
&=-n_\chi\sigmatwo \left(T_\nu \over 1.95~\text{K}\right)^2 \left[J(\mathbf{x},\hat{p},t) - {1 \over 4\pi}\int d^2\hat{p}'J(\mathbf{x},\hat{p}',t) \right].
\end{align}
We have used the following integral in the second last line,
\begin{equation}
{ \int_{0}^{\infty}  \bar{f}_\nu p^5 dp \over \int_{0}^{\infty} \bar{f}_\nu p^3 dp} = \frac{310 \pi ^2 T_\nu^2}{147}
\end{equation}
and defined
\begin{equation}
\sigmatwo \ \equiv \ \tilde{\sigma}\left({3 \over 2}\right)\left(310\pi^2 \over 147\right)\left(T_{\nu,0} \over m_\chi\right)^2
\end{equation}
where $ T_{\nu,0} $ is neutrino temperature today.

\bibliographystyle{unsrtads}
\bibliography{draftV5}

\begin{thebibliography}{10}

\bibitem{staro1979}
A.~A. {Starobinskii}.
\newblock {Spectrum of relict gravitational radiation and the early state of
  the universe}.
\newblock {\em Soviet Journal of Experimental and Theoretical Physics Letters},
  30:682, December 1979.
\newblock
  {\small[\href{http://adsabs.harvard.edu/abs/1979JETPL..30..682S}{ADS}]}.

\bibitem{rsv1982}
V.~A. {Rubakov}, M.~V. {Sazhin}, and A.~V. {Veryaskin}.
\newblock {Graviton creation in the inflationary universe and the grand
  unification scale}.
\newblock {\em Physics Letters B}, 115:189--192, September 1982.
\newblock \href {http://dx.doi.org/10.1016/0370-2693(82)90641-4}
  {\path{[DOI]}},
  {\small[\href{http://adsabs.harvard.edu/abs/1982PhLB..115..189R}{ADS}]}.

\bibitem{polnarev1985}
A.~G. {Polnarev}.
\newblock {Polarization and Anisotropy Induced in the Microwave Background by
  Cosmological Gravitational Waves}.
\newblock {\em Soviet Astronomy}, 29:607--613, December 1985.
\newblock
  {\small[\href{http://adsabs.harvard.edu/abs/1985SvA....29..607P}{ADS}]}.

\bibitem{seljak1997}
U.~{Seljak}.
\newblock {Measuring Polarization in the Cosmic Microwave Background}.
\newblock {\em ApJ}, 482:6--16, June 1997.
\newblock \href {http://arxiv.org/abs/astro-ph/9608131}
  {\path{arXiv:astro-ph/9608131}}, \href {http://dx.doi.org/10.1086/304123}
  {\path{[DOI]}},
  {\small[\href{http://adsabs.harvard.edu/abs/1997ApJ...482....6S}{ADS}]}.

\bibitem{sz1997}
U.~{Seljak} and M.~{Zaldarriaga}.
\newblock {Signature of Gravity Waves in the Polarization of the Microwave
  Background}.
\newblock {\em Physical Review Letters}, 78:2054--2057, March 1997.
\newblock \href {http://arxiv.org/abs/astro-ph/9609169}
  {\path{arXiv:astro-ph/9609169}}, \href
  {http://dx.doi.org/10.1103/PhysRevLett.78.2054} {\path{[DOI]}},
  {\small[\href{http://adsabs.harvard.edu/abs/1997PhRvL..78.2054S}{ADS}]}.

\bibitem{kks1997}
M.~{Kamionkowski}, A.~{Kosowsky}, and A.~{Stebbins}.
\newblock {Statistics of cosmic microwave background polarization}.
\newblock {\em Phys.Rev.D}, 55:7368--7388, June 1997.
\newblock \href {http://arxiv.org/abs/astro-ph/9611125}
  {\path{arXiv:astro-ph/9611125}}, \href
  {http://dx.doi.org/10.1103/PhysRevD.55.7368} {\path{[DOI]}},
  {\small[\href{http://adsabs.harvard.edu/abs/1997PhRvD..55.7368K}{ADS}]}.

\bibitem{bicep2014}
{P.~A.~R. {Ade} et al.}
\newblock {Detection of B-Mode Polarization at Degree Angular Scales by
  BICEP2}.
\newblock {\em Physical Review Letters}, 112(24):241101, June 2014.
\newblock \href {http://arxiv.org/abs/1403.3985} {\path{arXiv:1403.3985}},
  \href {http://dx.doi.org/10.1103/PhysRevLett.112.241101} {\path{[DOI]}},
  {\small[\href{http://adsabs.harvard.edu/abs/2014PhRvL.112x1101A}{ADS}]}.

\bibitem{bicep2016}
{BICEP2 Collaboration} and {Keck Array Collaboration}.
\newblock {Improved Constraints on Cosmology and Foregrounds from BICEP2 and
  Keck Array Cosmic Microwave Background Data with Inclusion of 95 GHz Band}.
\newblock {\em Physical Review Letters}, 116(3):031302, January 2016.
\newblock \href {http://arxiv.org/abs/1510.09217} {\path{arXiv:1510.09217}},
  \href {http://dx.doi.org/10.1103/PhysRevLett.116.031302} {\path{[DOI]}},
  {\small[\href{http://adsabs.harvard.edu/abs/2016PhRvL.116c1302B}{ADS}]}.

\bibitem{natural-inflation1}
Katherine Freese, Joshua~A. Frieman, and Angela~V. Olinto.
\newblock Natural inflation with pseudo nambu-goldstone bosons.
\newblock {\em Phys. Rev. Lett.}, 65:3233--3236, Dec 1990.
\newblock URL: \url{https://link.aps.org/doi/10.1103/PhysRevLett.65.3233},
  \href {http://dx.doi.org/10.1103/PhysRevLett.65.3233} {\path{[DOI]}}.

\bibitem{natural-inflation2}
Eva Silverstein and Alexander Westphal.
\newblock {Monodromy in the CMB: Gravity Waves and String Inflation}.
\newblock {\em Phys. Rev.}, D78:106003, 2008.
\newblock \href {http://arxiv.org/abs/0803.3085} {\path{arXiv:0803.3085}},
  \href {http://dx.doi.org/10.1103/PhysRevD.78.106003} {\path{[DOI]}}.

\bibitem{modgr1}
Alexei~A. Starobinsky.
\newblock {A New Type of Isotropic Cosmological Models Without Singularity}.
\newblock {\em Phys. Lett.}, 91B:99--102, 1980.
\newblock \href {http://dx.doi.org/10.1016/0370-2693(80)90670-X}
  {\path{[DOI]}}.

\bibitem{modgr2}
John Ellis, Dimitri~V. Nanopoulos, and Keith~A. Olive.
\newblock {No-Scale Supergravity Realization of the Starobinsky Model of
  Inflation}.
\newblock {\em Phys. Rev. Lett.}, 111:111301, 2013.
\newblock [Erratum: Phys. Rev. Lett.111,no.12,129902(2013)].
\newblock \href {http://arxiv.org/abs/1305.1247} {\path{arXiv:1305.1247}},
  \href {http://dx.doi.org/10.1103/PhysRevLett.111.129902,
  10.1103/PhysRevLett.111.111301} {\path{[DOI]}}.

\bibitem{inflation-model1}
A.D. Linde.
\newblock Chaotic inflation.
\newblock {\em Physics Letters B}, 129(3):177 -- 181, 1983.
\newblock URL:
  \url{http://www.sciencedirect.com/science/article/pii/0370269383908377},
  \href {http://dx.doi.org/https://doi.org/10.1016/0370-2693(83)90837-7}
  {\path{[DOI]}}.

\bibitem{inflation-model2}
Andrei~D. Linde.
\newblock {Hybrid inflation}.
\newblock {\em Phys. Rev.}, D49:748--754, 1994.
\newblock \href {http://arxiv.org/abs/astro-ph/9307002}
  {\path{arXiv:astro-ph/9307002}}, \href
  {http://dx.doi.org/10.1103/PhysRevD.49.748} {\path{[DOI]}}.

\bibitem{primo-mag1}
Michael~S. Turner and Lawrence~M. Widrow.
\newblock Inflation-produced, large-scale magnetic fields.
\newblock {\em Phys. Rev. D}, 37:2743--2754, May 1988.
\newblock URL: \url{https://link.aps.org/doi/10.1103/PhysRevD.37.2743}, \href
  {http://dx.doi.org/10.1103/PhysRevD.37.2743} {\path{[DOI]}}.

\bibitem{primo-mag2}
W.~Daniel Garretson, George~B. Field, and Sean~M. Carroll.
\newblock {Primordial magnetic fields from pseudoGoldstone bosons}.
\newblock {\em Phys. Rev.}, D46:5346--5351, 1992.
\newblock \href {http://arxiv.org/abs/hep-ph/9209238}
  {\path{arXiv:hep-ph/9209238}}, \href
  {http://dx.doi.org/10.1103/PhysRevD.46.5346} {\path{[DOI]}}.

\bibitem{primo-mag3}
M.~Gasperini, Massimo Giovannini, and G.~Veneziano.
\newblock {Primordial magnetic fields from string cosmology}.
\newblock {\em Phys. Rev. Lett.}, 75:3796--3799, 1995.
\newblock \href {http://arxiv.org/abs/hep-th/9504083}
  {\path{arXiv:hep-th/9504083}}, \href
  {http://dx.doi.org/10.1103/PhysRevLett.75.3796} {\path{[DOI]}}.

\bibitem{Weinberg:2003ur}
Steven Weinberg.
\newblock {Damping of tensor modes in cosmology}.
\newblock {\em Phys. Rev.}, D69:023503, 2004.
\newblock \href {http://arxiv.org/abs/astro-ph/0306304}
  {\path{arXiv:astro-ph/0306304}}, \href
  {http://dx.doi.org/10.1103/PhysRevD.69.023503} {\path{[DOI]}}.

\bibitem{bullet2004}
D.~{Clowe}, A.~{Gonzalez}, and M.~{Markevitch}.
\newblock {Weak-Lensing Mass Reconstruction of the Interacting Cluster 1E
  0657-558: Direct Evidence for the Existence of Dark Matter}.
\newblock {\em ApJ}, 604:596--603, April 2004.
\newblock \href {http://arxiv.org/abs/astro-ph/0312273}
  {\path{arXiv:astro-ph/0312273}}, \href {http://dx.doi.org/10.1086/381970}
  {\path{[DOI]}},
  {\small[\href{http://adsabs.harvard.edu/abs/2004ApJ...604..596C}{ADS}]}.

\bibitem{bullet2004_2}
M.~{Markevitch}, A.~H. {Gonzalez}, D.~{Clowe}, A.~{Vikhlinin}, W.~{Forman},
  C.~{Jones}, S.~{Murray}, and W.~{Tucker}.
\newblock {Direct Constraints on the Dark Matter Self-Interaction Cross Section
  from the Merging Galaxy Cluster 1E 0657-56}.
\newblock {\em ApJ}, 606:819--824, May 2004.
\newblock \href {http://arxiv.org/abs/astro-ph/0309303}
  {\path{arXiv:astro-ph/0309303}}, \href {http://dx.doi.org/10.1086/383178}
  {\path{[DOI]}},
  {\small[\href{http://adsabs.harvard.edu/abs/2004ApJ...606..819M}{ADS}]}.

\bibitem{planck2015}
{Planck Collaboration}, P.~A.~R. {Ade}, N.~{Aghanim}, M.~{Arnaud},
  M.~{Ashdown}, J.~{Aumont}, C.~{Baccigalupi}, A.~J. {Banday}, R.~B.
  {Barreiro}, J.~G. {Bartlett}, and et~al .
\newblock {Planck 2015 results. XIII. Cosmological parameters}.
\newblock {\em A\&A}, 594:A13, September 2016.
\newblock \href {http://arxiv.org/abs/1502.01589} {\path{arXiv:1502.01589}},
  \href {http://dx.doi.org/10.1051/0004-6361/201525830} {\path{[DOI]}},
  {\small[\href{http://adsabs.harvard.edu/abs/2016A%26A...594A..13P}{ADS}]}.

\bibitem{Agnese:2015nto}
R.~Agnese et~al.
\newblock {New Results from the Search for Low-Mass Weakly Interacting Massive
  Particles with the CDMS Low Ionization Threshold Experiment}.
\newblock {\em Phys. Rev. Lett.}, 116(7):071301, 2016.
\newblock \href {http://arxiv.org/abs/1509.02448} {\path{arXiv:1509.02448}},
  \href {http://dx.doi.org/10.1103/PhysRevLett.116.071301} {\path{[DOI]}}.

\bibitem{Akerib:2016vxi}
D.~S. Akerib et~al.
\newblock {Results from a search for dark matter in the complete LUX exposure}.
\newblock {\em Phys. Rev. Lett.}, 118(2):021303, 2017.
\newblock \href {http://arxiv.org/abs/1608.07648} {\path{arXiv:1608.07648}},
  \href {http://dx.doi.org/10.1103/PhysRevLett.118.021303} {\path{[DOI]}}.

\bibitem{Aprile:2017iyp}
E.~Aprile et~al.
\newblock {First Dark Matter Search Results from the XENON1T Experiment}.
\newblock {\em Phys. Rev. Lett.}, 119(18):181301, 2017.
\newblock \href {http://arxiv.org/abs/1705.06655} {\path{arXiv:1705.06655}},
  \href {http://dx.doi.org/10.1103/PhysRevLett.119.181301} {\path{[DOI]}}.

\bibitem{Cui:2017nnn}
Xiangyi Cui et~al.
\newblock {Dark Matter Results From 54-Ton-Day Exposure of PandaX-II
  Experiment}.
\newblock {\em Phys. Rev. Lett.}, 119(18):181302, 2017.
\newblock \href {http://arxiv.org/abs/1708.06917} {\path{arXiv:1708.06917}},
  \href {http://dx.doi.org/10.1103/PhysRevLett.119.181302} {\path{[DOI]}}.

\bibitem{2010PhRvL.104d1301A}
S.~J. {Asztalos}, G.~{Carosi}, C.~{Hagmann}, D.~{Kinion}, K.~{van Bibber},
  M.~{Hotz}, L.~J. {Rosenberg}, G.~{Rybka}, J.~{Hoskins}, J.~{Hwang},
  P.~{Sikivie}, D.~B. {Tanner}, R.~{Bradley}, J.~{Clarke}, and {ADMX
  Collaboration}.
\newblock {SQUID-Based Microwave Cavity Search for Dark-Matter Axions}.
\newblock {\em Physical Review Letters}, 104(4):041301, January 2010.
\newblock \href {http://arxiv.org/abs/0910.5914} {\path{arXiv:0910.5914}},
  \href {http://dx.doi.org/10.1103/PhysRevLett.104.041301} {\path{[DOI]}},
  {\small[\href{http://adsabs.harvard.edu/abs/2010PhRvL.104d1301A}{ADS}]}.

\bibitem{Arik:2015cjv}
M.~Arik et~al.
\newblock {New solar axion search using the CERN Axion Solar Telescope with
  $^4$He filling}.
\newblock {\em Phys. Rev.}, D92(2):021101, 2015.
\newblock \href {http://arxiv.org/abs/1503.00610} {\path{arXiv:1503.00610}},
  \href {http://dx.doi.org/10.1103/PhysRevD.92.021101} {\path{[DOI]}}.

\bibitem{Vogel:2013bta}
J.~K. Vogel et~al.
\newblock {IAXO - The International Axion Observatory}.
\newblock In {\em {8th Patras Workshop on Axions, WIMPs and WISPs (AXION-WIMP
  2012) Chicago, Illinois, July 18-22, 2012}}, 2013.
\newblock URL:
  \url{http://lss.fnal.gov/archive/2013/pub/fermilab-pub-13-699-a.pdf}, \href
  {http://arxiv.org/abs/1302.3273} {\path{arXiv:1302.3273}}.

\bibitem{Budker:2013hfa}
Dmitry Budker, Peter~W. Graham, Micah Ledbetter, Surjeet Rajendran, and Alex
  Sushkov.
\newblock {Proposal for a Cosmic Axion Spin Precession Experiment (CASPEr)}.
\newblock {\em Phys. Rev.}, X4(2):021030, 2014.
\newblock \href {http://arxiv.org/abs/1306.6089} {\path{arXiv:1306.6089}},
  \href {http://dx.doi.org/10.1103/PhysRevX.4.021030} {\path{[DOI]}}.

\bibitem{Abel:2017rtm}
C.~Abel et~al.
\newblock {Search for Axionlike Dark Matter through Nuclear Spin Precession in
  Electric and Magnetic Fields}.
\newblock {\em Phys. Rev.}, X7(4):041034, 2017.
\newblock \href {http://arxiv.org/abs/1708.06367} {\path{arXiv:1708.06367}},
  \href {http://dx.doi.org/10.1103/PhysRevX.7.041034} {\path{[DOI]}}.

\bibitem{TheFermi-LAT:2017vmf}
M.~Ackermann et~al.
\newblock {The Fermi Galactic Center GeV Excess and Implications for Dark
  Matter}.
\newblock {\em Astrophys. J.}, 840(1):43, 2017.
\newblock \href {http://arxiv.org/abs/1704.03910} {\path{arXiv:1704.03910}},
  \href {http://dx.doi.org/10.3847/1538-4357/aa6cab} {\path{[DOI]}}.

\bibitem{Abdallah:2016ygi}
H.~Abdallah et~al.
\newblock {Search for dark matter annihilations towards the inner Galactic halo
  from 10 years of observations with H.E.S.S}.
\newblock {\em Phys. Rev. Lett.}, 117(11):111301, 2016.
\newblock \href {http://arxiv.org/abs/1607.08142} {\path{arXiv:1607.08142}},
  \href {http://dx.doi.org/10.1103/PhysRevLett.117.111301} {\path{[DOI]}}.

\bibitem{Ahnen:2016qkx}
M.~L. Ahnen et~al.
\newblock {Limits to dark matter annihilation cross-section from a combined
  analysis of MAGIC and Fermi-LAT observations of dwarf satellite galaxies}.
\newblock {\em JCAP}, 1602(02):039, 2016.
\newblock \href {http://arxiv.org/abs/1601.06590} {\path{arXiv:1601.06590}},
  \href {http://dx.doi.org/10.1088/1475-7516/2016/02/039} {\path{[DOI]}}.

\bibitem{Collaboration:2011jza}
A.~Gando et~al.
\newblock {A study of extraterrestrial antineutrino sources with the KamLAND
  detector}.
\newblock {\em Astrophys. J.}, 745:193, 2012.
\newblock \href {http://arxiv.org/abs/1105.3516} {\path{arXiv:1105.3516}},
  \href {http://dx.doi.org/10.1088/0004-637X/745/2/193} {\path{[DOI]}}.

\bibitem{Frankiewicz:2015zma}
Katarzyna Frankiewicz.
\newblock {Searching for Dark Matter Annihilation into Neutrinos with
  Super-Kamiokande}.
\newblock In {\em {Proceedings, Meeting of the APS Division of Particles and
  Fields (DPF 2015): Ann Arbor, Michigan, USA, 4-8 Aug 2015}}, 2015.
\newblock URL:
  \url{https://inspirehep.net/record/1401009/files/arXiv:1510.07999.pdf}, \href
  {http://arxiv.org/abs/1510.07999} {\path{arXiv:1510.07999}}.

\bibitem{icecube2017}
{IceCube Collaboration}, M.~G. {Aartsen}, M.~{Ackermann}, J.~{Adams}, J.~A.
  {Aguilar}, M.~{Ahlers}, M.~{Ahrens}, I.~A. {Samarai}, D.~{Altmann},
  K.~{Andeen}, and et~al.
\newblock {Search for Neutrinos from Dark Matter Self-Annihilations in the
  center of the Milky Way with 3 years of IceCube/DeepCore}.
\newblock {\em ArXiv e-prints}, May 2017.
\newblock \href {http://arxiv.org/abs/1705.08103} {\path{arXiv:1705.08103}},
  {\small[\href{http://adsabs.harvard.edu/abs/2017arXiv170508103I}{ADS}]}.

\bibitem{Primulando:2017kxf}
Reinard Primulando and Patipan Uttayarat.
\newblock {Dark Matter-Neutrino Interaction in Light of Collider and Neutrino
  Telescope Data}.
\newblock 2017.
\newblock \href {http://arxiv.org/abs/1710.08567} {\path{arXiv:1710.08567}}.

\bibitem{Moore:1994yx}
B.~Moore.
\newblock {Evidence against dissipationless dark matter from observations of
  galaxy haloes}.
\newblock {\em Nature}, 370:629, 1994.
\newblock \href {http://dx.doi.org/10.1038/370629a0} {\path{[DOI]}}.

\bibitem{Kravtsov:1997dp}
Andrey~V. Kravtsov, Anatoly~A. Klypin, James~S. Bullock, and Joel~R. Primack.
\newblock {The Cores of dark matter dominated galaxies: Theory versus
  observations}.
\newblock {\em Astrophys. J.}, 502:48, 1998.
\newblock \href {http://arxiv.org/abs/astro-ph/9708176}
  {\path{arXiv:astro-ph/9708176}}, \href {http://dx.doi.org/10.1086/305884}
  {\path{[DOI]}}.

\bibitem{Boehm:2000gq}
C.~Boehm, Pierre Fayet, and R.~Schaeffer.
\newblock {Constraining dark matter candidates from structure formation}.
\newblock {\em Phys. Lett.}, B518:8--14, 2001.
\newblock \href {http://arxiv.org/abs/astro-ph/0012504}
  {\path{arXiv:astro-ph/0012504}}, \href
  {http://dx.doi.org/10.1016/S0370-2693(01)01060-7} {\path{[DOI]}}.

\bibitem{ah2014}
M.~{Archidiacono} and S.~{Hannestad}.
\newblock {Updated constraints on non-standard neutrino interactions from
  Planck}.
\newblock {\em JCAP}, 7:046, July 2014.
\newblock \href {http://arxiv.org/abs/1311.3873} {\path{arXiv:1311.3873}},
  \href {http://dx.doi.org/10.1088/1475-7516/2014/07/046} {\path{[DOI]}},
  {\small[\href{http://adsabs.harvard.edu/abs/2014JCAP...07..046A}{ADS}]}.

\bibitem{kt2014}
P.~{Ko} and Y.~{Tang}.
\newblock {{$\nu$}{$\Lambda$}MDM: A model for sterile neutrino and dark matter
  reconciles cosmological and neutrino oscillation data after BICEP2}.
\newblock {\em Physics Letters B}, 739:62--67, December 2014.
\newblock \href {http://arxiv.org/abs/1404.0236} {\path{arXiv:1404.0236}},
  \href {http://dx.doi.org/10.1016/j.physletb.2014.10.035} {\path{[DOI]}},
  {\small[\href{http://adsabs.harvard.edu/abs/2014PhLB..739...62K}{ADS}]}.

\bibitem{Wilkinson:2014ksa}
Ryan~J. Wilkinson, Celine Boehm, and Julien Lesgourgues.
\newblock {Constraining Dark Matter-Neutrino Interactions using the CMB and
  Large-Scale Structure}.
\newblock {\em JCAP}, 1405:011, 2014.
\newblock \href {http://arxiv.org/abs/1401.7597} {\path{arXiv:1401.7597}},
  \href {http://dx.doi.org/10.1088/1475-7516/2014/05/011} {\path{[DOI]}}.

\bibitem{Bertoni:2014mva}
Bridget Bertoni, Seyda Ipek, David McKeen, and Ann~E. Nelson.
\newblock {Constraints and consequences of reducing small scale structure via
  large dark matter-neutrino interactions}.
\newblock {\em JHEP}, 04:170, 2015.
\newblock \href {http://arxiv.org/abs/1412.3113} {\path{arXiv:1412.3113}},
  \href {http://dx.doi.org/10.1007/JHEP04(2015)170} {\path{[DOI]}}.

\bibitem{Boehm:2014vja}
C.~Boehm, J.~A. Schewtschenko, R.~J. Wilkinson, C.~M. Baugh, and S.~Pascoli.
\newblock {Using the Milky Way satellites to study interactions between cold
  dark matter and radiation}.
\newblock {\em Mon. Not. Roy. Astron. Soc.}, 445:L31--L35, 2014.
\newblock \href {http://arxiv.org/abs/1404.7012} {\path{arXiv:1404.7012}},
  \href {http://dx.doi.org/10.1093/mnrasl/slu115} {\path{[DOI]}}.

\bibitem{abc2009}
L.~{Ackerman}, M.~R. {Buckley}, S.~M. {Carroll}, and M.~{Kamionkowski}.
\newblock {Dark matter and dark radiation}.
\newblock {\em Phys.Rev.D}, 79(2):023519, January 2009.
\newblock \href {http://arxiv.org/abs/0810.5126} {\path{arXiv:0810.5126}},
  \href {http://dx.doi.org/10.1103/PhysRevD.79.023519} {\path{[DOI]}},
  {\small[\href{http://adsabs.harvard.edu/abs/2009PhRvD..79b3519A}{ADS}]}.

\bibitem{fky2010}
J.~L. {Feng}, M.~{Kaplinghat}, and H.-B. {Yu}.
\newblock {Sommerfeld enhancements for thermal relic dark matter}.
\newblock {\em Phys.Rev.D}, 82(8):083525, October 2010.
\newblock \href {http://arxiv.org/abs/1005.4678} {\path{arXiv:1005.4678}},
  \href {http://dx.doi.org/10.1103/PhysRevD.82.083525} {\path{[DOI]}},
  {\small[\href{http://adsabs.harvard.edu/abs/2010PhRvD..82h3525F}{ADS}]}.

\bibitem{bfm2012}
M.~{Blennow}, E.~{Fernandez Martinez}, O.~{Mena}, J.~{Redondo}, and e.~P.
  {Serra}.
\newblock {Asymmetric Dark Matter and Dark Radiation}.
\newblock {\em JCAP}, 7:022, July 2012.
\newblock \href {http://arxiv.org/abs/1203.5803} {\path{arXiv:1203.5803}},
  \href {http://dx.doi.org/10.1088/1475-7516/2012/07/022} {\path{[DOI]}},
  {\small[\href{http://adsabs.harvard.edu/abs/2012JCAP...07..022B}{ADS}]}.

\bibitem{cpr2014}
F.-Y. {Cyr-Racine}, R.~{de Putter}, A.~{Raccanelli}, and K.~{Sigurdson}.
\newblock {Constraints on large-scale dark acoustic oscillations from
  cosmology}.
\newblock {\em Phys.Rev.D}, 89(6):063517, March 2014.
\newblock \href {http://arxiv.org/abs/1310.3278} {\path{arXiv:1310.3278}},
  \href {http://dx.doi.org/10.1103/PhysRevD.89.063517} {\path{[DOI]}},
  {\small[\href{http://adsabs.harvard.edu/abs/2014PhRvD..89f3517C}{ADS}]}.

\bibitem{cd2014}
X.~{Chu} and B.~{Dasgupta}.
\newblock {Dark Radiation Alleviates Problems with Dark Matter Halos}.
\newblock {\em Physical Review Letters}, 113(16):161301, October 2014.
\newblock \href {http://arxiv.org/abs/1404.6127} {\path{arXiv:1404.6127}},
  \href {http://dx.doi.org/10.1103/PhysRevLett.113.161301} {\path{[DOI]}},
  {\small[\href{http://adsabs.harvard.edu/abs/2014PhRvL.113p1301C}{ADS}]}.

\bibitem{bsl2017}
M.~A. {Buen-Abad}, M.~{Schmaltz}, J.~{Lesgourgues}, and T.~{Brinckmann}.
\newblock {Interacting Dark Sector and Precision Cosmology}.
\newblock {\em ArXiv e-prints}, August 2017.
\newblock \href {http://arxiv.org/abs/1708.09406} {\path{arXiv:1708.09406}},
  {\small[\href{http://adsabs.harvard.edu/abs/2017arXiv170809406B}{ADS}]}.

\bibitem{Spergel:1999mh}
David~N. Spergel and Paul~J. Steinhardt.
\newblock {Observational evidence for selfinteracting cold dark matter}.
\newblock {\em Phys. Rev. Lett.}, 84:3760--3763, 2000.
\newblock \href {http://arxiv.org/abs/astro-ph/9909386}
  {\path{arXiv:astro-ph/9909386}}, \href
  {http://dx.doi.org/10.1103/PhysRevLett.84.3760} {\path{[DOI]}}.

\bibitem{Foot:2014uba}
R.~Foot and S.~Vagnozzi.
\newblock {Dissipative hidden sector dark matter}.
\newblock {\em Phys. Rev.}, D91:023512, 2015.
\newblock \href {http://arxiv.org/abs/1409.7174} {\path{arXiv:1409.7174}},
  \href {http://dx.doi.org/10.1103/PhysRevD.91.023512} {\path{[DOI]}}.

\bibitem{Chacko:2015noa}
Zackaria Chacko, Yanou Cui, Sungwoo Hong, and Takemichi Okui.
\newblock {Hidden dark matter sector, dark radiation, and the CMB}.
\newblock {\em Phys. Rev.}, D92:055033, 2015.
\newblock \href {http://arxiv.org/abs/1505.04192} {\path{arXiv:1505.04192}},
  \href {http://dx.doi.org/10.1103/PhysRevD.92.055033} {\path{[DOI]}}.

\bibitem{Belotsky:2001fb}
K.~M. Belotsky, A.~L. Sudarikov, and M.~{\relax Yu}. Khlopov.
\newblock {Constraint on anomalous 4nu interaction}.
\newblock {\em Phys. Atom. Nucl.}, 64:1637--1642, 2001.
\newblock [Yad. Fiz.64,1718(2001)].
\newblock \href {http://dx.doi.org/10.1134/1.1409505} {\path{[DOI]}}.

\bibitem{Berkov:1987pz}
A.~V. Berkov, {\relax Yu}.~P. Nikitin, A.~L. Sudarikov, and M.~{\relax Yu}.
  Khlopov.
\newblock {POSSIBLE EXPERIMENTAL SEARCH FOR ANOMALOUS 4 NEUTRINO INTERACTION.
  (IN RUSSIAN)}.
\newblock {\em Yad. Fiz.}, 46:1729--1737, 1987.

\bibitem{1983ApJ...274L...1W}
S.~D.~M. {White}, C.~S. {Frenk}, and M.~{Davis}.
\newblock {Clustering in a neutrino-dominated universe}.
\newblock {\em ApJL}, 274:L1--L5, November 1983.
\newblock \href {http://dx.doi.org/10.1086/184139} {\path{[DOI]}},
  {\small[\href{http://adsabs.harvard.edu/abs/1983ApJ...274L...1W}{ADS}]}.

\bibitem{doi:10.1143/PTPS.78.1}
Hideo Kodama and Misao Sasaki.
\newblock Cosmological perturbation theory.
\newblock {\em Progress of Theoretical Physics Supplement}, 78:1--166, 1984.
\newblock URL: \url{+ http://dx.doi.org/10.1143/PTPS.78.1}, \href
  {http://arxiv.org/abs//oup/backfile/content_public/journal/ptps/78/10.1143/ptps.78.1/2/78-1.pdf}
  {\path{arXiv:/oup/backfile/content_public/journal/ptps/78/10.1143/ptps.78.1/2/78-1.pdf}},
  \href {http://dx.doi.org/10.1143/PTPS.78.1} {\path{[DOI]}}.

\bibitem{zks1968}
Y.~B. {Zeldovich}, V.~G. {Kurt}, and R.~A. {Sunyaev}.
\newblock {Recombination of Hydrogen in the Hot Model of the Universe}.
\newblock {\em Zhurnal Eksperimentalnoi i Teoreticheskoi Fiziki}, 55:278--286,
  July 1968.
\newblock
  {\small[\href{http://adsabs.harvard.edu/abs/1968ZhETF..55..278Z}{ADS}]}.

\bibitem{peebles1968}
P.~J.~E. {Peebles}.
\newblock {Recombination of the Primeval Plasma}.
\newblock {\em \apj}, 153:1, July 1968.
\newblock \href {http://dx.doi.org/10.1086/149628} {\path{[DOI]}},
  {\small[\href{http://adsabs.harvard.edu/abs/1968ApJ...153....1P}{ADS}]}.

\bibitem{cs2012}
J.~{Chluba} and R.~A. {Sunyaev}.
\newblock {The evolution of CMB spectral distortions in the early Universe}.
\newblock {\em \mnras}, 419:1294--1314, January 2012.
\newblock \href {http://arxiv.org/abs/1109.6552} {\path{arXiv:1109.6552}},
  \href {http://dx.doi.org/10.1111/j.1365-2966.2011.19786.x} {\path{[DOI]}},
  {\small[\href{http://adsabs.harvard.edu/abs/2012MNRAS.419.1294C}{ADS}]}.

\bibitem{ksc2012}
R.~{Khatri}, R.~A. {Sunyaev}, and J.~{Chluba}.
\newblock {Does Bose-Einstein condensation of CMB photons cancel {$\mu$}
  distortions created by dissipation of sound waves in the early Universe?}
\newblock {\em \aap}, 540:A124, April 2012.
\newblock \href {http://arxiv.org/abs/1110.0475} {\path{arXiv:1110.0475}},
  \href {http://dx.doi.org/10.1051/0004-6361/201118194} {\path{[DOI]}},
  {\small[\href{http://adsabs.harvard.edu/abs/2012A%26A...540A.124K}{ADS}]}.

\bibitem{weinberg}
S.~{Weinberg}.
\newblock {\em Cosmology}.
\newblock Oxford University Press, Oxford, 2008.

\bibitem{Ma:1995ey}
Chung-Pei Ma and Edmund Bertschinger.
\newblock {Cosmological perturbation theory in the synchronous and conformal
  Newtonian gauges}.
\newblock {\em Astrophys. J.}, 455:7--25, 1995.
\newblock \href {http://arxiv.org/abs/astro-ph/9506072}
  {\path{arXiv:astro-ph/9506072}}, \href {http://dx.doi.org/10.1086/176550}
  {\path{[DOI]}}.

\bibitem{2011JCAP...07..034B}
D.~{Blas}, J.~{Lesgourgues}, and T.~{Tram}.
\newblock {The Cosmic Linear Anisotropy Solving System (CLASS). Part II:
  Approximation schemes}.
\newblock {\em JCAP}, 7:034, July 2011.
\newblock \href {http://arxiv.org/abs/1104.2933} {\path{arXiv:1104.2933}},
  \href {http://dx.doi.org/10.1088/1475-7516/2011/07/034} {\path{[DOI]}},
  {\small[\href{http://adsabs.harvard.edu/abs/2011JCAP...07..034B}{ADS}]}.

\bibitem{Goldberg1967}
J.~N. {Goldberg}, A.~J. {Macfarlane}, E.~T. {Newman}, F.~{Rohrlich}, and
  E.~C.~G. {Sudarshan}.
\newblock {Spin-s Spherical Harmonics and {$\eth$}}.
\newblock {\em Journal of Mathematical Physics}, 8:2155--2161, November 1967.
\newblock \href {http://dx.doi.org/10.1063/1.1705135} {\path{[DOI]}},
  {\small[\href{http://adsabs.harvard.edu/abs/1967JMP.....8.2155G}{ADS}]}.

\bibitem{knox2002}
L.~{Knox} and Y.-S. {Song}.
\newblock {Limit on the Detectability of the Energy Scale of Inflation}.
\newblock {\em Physical Review Letters}, 89(1):011303, July 2002.
\newblock \href {http://arxiv.org/abs/astro-ph/0202286}
  {\path{arXiv:astro-ph/0202286}}, \href
  {http://dx.doi.org/10.1103/PhysRevLett.89.011303} {\path{[DOI]}},
  {\small[\href{http://adsabs.harvard.edu/abs/2002PhRvL..89a1303K}{ADS}]}.

\bibitem{kesden2002}
M.~{Kesden}, A.~{Cooray}, and M.~{Kamionkowski}.
\newblock {Separation of Gravitational-Wave and Cosmic-Shear Contributions to
  Cosmic Microwave Background Polarization}.
\newblock {\em Physical Review Letters}, 89(1):011304, July 2002.
\newblock \href {http://arxiv.org/abs/astro-ph/0202434}
  {\path{arXiv:astro-ph/0202434}}, \href
  {http://dx.doi.org/10.1103/PhysRevLett.89.011304} {\path{[DOI]}},
  {\small[\href{http://adsabs.harvard.edu/abs/2002PhRvL..89a1304K}{ADS}]}.

\bibitem{seljak2004}
U.~{Seljak} and C.~M. {Hirata}.
\newblock {Gravitational lensing as a contaminant of the gravity wave signal in
  the CMB}.
\newblock {\em Phys.Rev.D}, 69(4):043005, February 2004.
\newblock \href {http://arxiv.org/abs/astro-ph/0310163}
  {\path{arXiv:astro-ph/0310163}}, \href
  {http://dx.doi.org/10.1103/PhysRevD.69.043005} {\path{[DOI]}},
  {\small[\href{http://adsabs.harvard.edu/abs/2004PhRvD..69d3005S}{ADS}]}.

\bibitem{marian2007}
L.~{Marian} and G.~M. {Bernstein}.
\newblock {Detectability of CMB tensor B modes via delensing with weak lensing
  galaxy surveys}.
\newblock {\em Phys.Rev.D}, 76(12):123009, December 2007.
\newblock \href {http://arxiv.org/abs/0710.2538} {\path{arXiv:0710.2538}},
  \href {http://dx.doi.org/10.1103/PhysRevD.76.123009} {\path{[DOI]}},
  {\small[\href{http://adsabs.harvard.edu/abs/2007PhRvD..76l3009M}{ADS}]}.

\bibitem{smith2012}
K.~M. {Smith}, D.~{Hanson}, M.~{LoVerde}, C.~M. {Hirata}, and O.~{Zahn}.
\newblock {Delensing CMB polarization with external datasets}.
\newblock {\em JCAP}, 6:014, June 2012.
\newblock \href {http://arxiv.org/abs/1010.0048} {\path{arXiv:1010.0048}},
  \href {http://dx.doi.org/10.1088/1475-7516/2012/06/014} {\path{[DOI]}},
  {\small[\href{http://adsabs.harvard.edu/abs/2012JCAP...06..014S}{ADS}]}.

\bibitem{simrad2015}
G.~{Simard}, D.~{Hanson}, and G.~{Holder}.
\newblock {Prospects for Delensing the Cosmic Microwave Background for Studying
  Inflation}.
\newblock {\em ApJ}, 807:166, July 2015.
\newblock \href {http://arxiv.org/abs/1410.0691} {\path{arXiv:1410.0691}},
  \href {http://dx.doi.org/10.1088/0004-637X/807/2/166} {\path{[DOI]}},
  {\small[\href{http://adsabs.harvard.edu/abs/2015ApJ...807..166S}{ADS}]}.

\bibitem{sherwin2015}
B.~D. {Sherwin} and M.~{Schmittfull}.
\newblock {Delensing the CMB with the cosmic infrared background}.
\newblock {\em Phys.Rev.D}, 92(4):043005, August 2015.
\newblock \href {http://arxiv.org/abs/1502.05356} {\path{arXiv:1502.05356}},
  \href {http://dx.doi.org/10.1103/PhysRevD.92.043005} {\path{[DOI]}},
  {\small[\href{http://adsabs.harvard.edu/abs/2015PhRvD..92d3005S}{ADS}]}.

\bibitem{namikawa2016}
T.~{Namikawa}, D.~{Yamauchi}, B.~{Sherwin}, and R.~{Nagata}.
\newblock {Delensing cosmic microwave background B modes with the Square
  Kilometre Array Radio Continuum Survey}.
\newblock {\em Phys.Rev.D}, 93(4):043527, February 2016.
\newblock \href {http://arxiv.org/abs/1511.04653} {\path{arXiv:1511.04653}},
  \href {http://dx.doi.org/10.1103/PhysRevD.93.043527} {\path{[DOI]}},
  {\small[\href{http://adsabs.harvard.edu/abs/2016PhRvD..93d3527N}{ADS}]}.

\bibitem{k2016}
M.~{Kamionkowski} and E.~D. {Kovetz}.
\newblock {The Quest for B Modes from Inflationary Gravitational Waves}.
\newblock {\em ARA\&A}, 54:227--269, September 2016.
\newblock \href {http://arxiv.org/abs/1510.06042} {\path{arXiv:1510.06042}},
  \href {http://dx.doi.org/10.1146/annurev-astro-081915-023433} {\path{[DOI]}},
  {\small[\href{http://adsabs.harvard.edu/abs/2016ARA%26A..54..227K}{ADS}]}.

\bibitem{larsen2016}
P.~{Larsen}, A.~{Challinor}, B.~D. {Sherwin}, and D.~{Mak}.
\newblock {Demonstration of Cosmic Microwave Background Delensing Using the
  Cosmic Infrared Background}.
\newblock {\em Physical Review Letters}, 117(15):151102, October 2016.
\newblock \href {http://arxiv.org/abs/1607.05733} {\path{arXiv:1607.05733}},
  \href {http://dx.doi.org/10.1103/PhysRevLett.117.151102} {\path{[DOI]}},
  {\small[\href{http://adsabs.harvard.edu/abs/2016PhRvL.117o1102L}{ADS}]}.

\bibitem{carron2017}
J.~{Carron}, A.~{Lewis}, and A.~{Challinor}.
\newblock {Internal delensing of Planck CMB temperature and polarization}.
\newblock {\em JCAP}, 5:035, May 2017.
\newblock \href {http://arxiv.org/abs/1701.01712} {\path{arXiv:1701.01712}},
  \href {http://dx.doi.org/10.1088/1475-7516/2017/05/035} {\path{[DOI]}},
  {\small[\href{http://adsabs.harvard.edu/abs/2017JCAP...05..035C}{ADS}]}.

\bibitem{sehgal2017}
N.~{Sehgal}, M.~S. {Madhavacheril}, B.~{Sherwin}, and A.~{van Engelen}.
\newblock {Internal delensing of cosmic microwave background acoustic peaks}.
\newblock {\em Phys.Rev.D}, 95(10):103512, May 2017.
\newblock \href {http://arxiv.org/abs/1612.03898} {\path{arXiv:1612.03898}},
  \href {http://dx.doi.org/10.1103/PhysRevD.95.103512} {\path{[DOI]}},
  {\small[\href{http://adsabs.harvard.edu/abs/2017PhRvD..95j3512S}{ADS}]}.

\bibitem{namikawa2017}
T.~{Namikawa}.
\newblock {CMB internal delensing with general optimal estimator for
  higher-order correlations}.
\newblock {\em Phys.Rev.D}, 95(10):103514, May 2017.
\newblock \href {http://arxiv.org/abs/1703.00169} {\path{arXiv:1703.00169}},
  \href {http://dx.doi.org/10.1103/PhysRevD.95.103514} {\path{[DOI]}},
  {\small[\href{http://adsabs.harvard.edu/abs/2017PhRvD..95j3514N}{ADS}]}.

\bibitem{rem2017}
M.~{Remazeilles}, C.~{Dickinson}, H.~K. {Eriksen}, and I.~K. {Wehus}.
\newblock {Joint Bayesian estimation of tensor and lensing B-modes in the power
  spectrum of CMB polarization data}.
\newblock {\em ArXiv e-prints}, July 2017.
\newblock \href {http://arxiv.org/abs/1707.02981} {\path{arXiv:1707.02981}},
  {\small[\href{http://adsabs.harvard.edu/abs/2017arXiv170702981R}{ADS}]}.

\bibitem{manzotti2017}
A.~{Manzotti}.
\newblock {Future cosmic microwave background delensing with galaxy surveys}.
\newblock {\em ArXiv e-prints}, October 2017.
\newblock \href {http://arxiv.org/abs/1710.11038} {\path{arXiv:1710.11038}},
  {\small[\href{http://adsabs.harvard.edu/abs/2017arXiv171011038M}{ADS}]}.

\bibitem{Escudero:2015yka}
Miguel Escudero, Olga Mena, Aaron~C. Vincent, Ryan~J. Wilkinson, and Céline
  Bœhm.
\newblock {Exploring dark matter microphysics with galaxy surveys}.
\newblock {\em JCAP}, 1509(09):034, 2015.
\newblock \href {http://arxiv.org/abs/1505.06735} {\path{arXiv:1505.06735}},
  \href {http://dx.doi.org/10.1088/1475-7516/2015/9/034,
  10.1088/1475-7516/2015/09/034} {\path{[DOI]}}.

\bibitem{dival2017}
E.~{Di Valentino}, C.~{B{\o}ehm}, E.~{Hivon}, and F.~R. {Bouchet}.
\newblock {Reducing the $H\_0$ and $\sigma\_8$ tensions with Dark
  Matter-neutrino interactions}.
\newblock {\em ArXiv e-prints}, October 2017.
\newblock \href {http://arxiv.org/abs/1710.02559} {\path{arXiv:1710.02559}},
  {\small[\href{http://adsabs.harvard.edu/abs/2017arXiv171002559D}{ADS}]}.

\bibitem{2010PhRvD..81d3507S}
P.~{Serra}, F.~{Zalamea}, A.~{Cooray}, G.~{Mangano}, and A.~{Melchiorri}.
\newblock {Constraints on neutrino-dark matter interactions from cosmic
  microwave background and large scale structure data}.
\newblock {\em Phys.Rev.D}, 81(4):043507, February 2010.
\newblock \href {http://arxiv.org/abs/0911.4411} {\path{arXiv:0911.4411}},
  \href {http://dx.doi.org/10.1103/PhysRevD.81.043507} {\path{[DOI]}},
  {\small[\href{http://adsabs.harvard.edu/abs/2010PhRvD..81d3507S}{ADS}]}.

\bibitem{futurepaper}
S.~{Ghosh}, R.~{Khatri}, and T.~{Roy}.
\newblock {\em in prep}, 2017.

\bibitem{bond1996}
J.~R. {Bond}.
\newblock {Theory and Observations of the Cosmic Background Radiation}.
\newblock In R.~{Schaeffer}, J.~{Silk}, M.~{Spiro}, and J.~{Zinn-Justin},
  editors, {\em Cosmology and Large Scale Structure}, page 469, January 1996.
\newblock
  {\small[\href{http://adsabs.harvard.edu/abs/1996clss.conf..469B}{ADS}]}.

\bibitem{cmbs4}
{K. N. Abazajian et al.}
\newblock {CMB-S4 Science Book, First Edition}.
\newblock {\em ArXiv e-prints}, October 2016.
\newblock \href {http://arxiv.org/abs/1610.02743} {\path{arXiv:1610.02743}},
  {\small[\href{http://adsabs.harvard.edu/abs/2016arXiv161002743A}{ADS}]}.

\bibitem{core2016}
{CORE Collaboration}.
\newblock {Exploring Cosmic Origins with CORE: Inflation}.
\newblock {\em ArXiv e-prints}, December 2016.
\newblock \href {http://arxiv.org/abs/1612.08270} {\path{arXiv:1612.08270}},
  {\small[\href{http://adsabs.harvard.edu/abs/2016arXiv161208270C}{ADS}]}.

\bibitem{prism}
{P. {Andr{\'e}} et al.}
\newblock {PRISM (Polarized Radiation Imaging and Spectroscopy Mission): an
  extended white paper}.
\newblock {\em JCAP}, 2:006, February 2014.
\newblock \href {http://arxiv.org/abs/1310.1554} {\path{arXiv:1310.1554}},
  \href {http://dx.doi.org/10.1088/1475-7516/2014/02/006} {\path{[DOI]}},
  {\small[\href{http://adsabs.harvard.edu/abs/2014JCAP...02..006A}{ADS}]}.

\bibitem{Diacoumis:2017hff}
James A.~D. Diacoumis and Yvonne Y.~Y. Wong.
\newblock {Using CMB spectral distortions to distinguish between dark matter
  solutions to the small-scale crisis}.
\newblock {\em JCAP}, 1709(09):011, 2017.
\newblock \href {http://arxiv.org/abs/1707.07050} {\path{arXiv:1707.07050}},
  \href {http://dx.doi.org/10.1088/1475-7516/2017/09/011} {\path{[DOI]}}.

\end{thebibliography}

\end{document}